# Massive Extremely High-Velocity Outflow in the Quasar J164653.72+243942.2

Paola Rodríguez Hidalgo,[1] Hyunseop Choi (최현섭),[2,3] Patrick B. Hall,[4] Karen M. Leighly,[5]
Liliana Flores,[1] Mikel M. Charles,[1,6] Cora DeFrancesco,[5] Julie Hlavacek-Larrondo,[2] and
Laurence Perreault-Levasseur[2,3]

[1]*Physical Sciences Division, School of STEM, University of Washington Bothell, WA, 98011, USA*
[2]*Département de Physique, Université de Montréal, Succ. Centre-Ville, Montréal, QC, H3C 3J7, Canada*
[3]*Mila - Quebec Artificial Intelligence Institute, Montréal, QC, H2S 3H1, Canada*
[4]*Department of Physics and Astronomy, York University, Toronto, ON, M3J 1P3, Canada*
[5]*Homer L. Dodge Department of Physics and Astronomy, The University of Oklahoma, Norman, OK, 73019, USA*
[6]*Department of Physics, Ohio State University, Columbus, OH, 43210, USA*

## ABSTRACT

We present the analysis of one of the most extreme quasar outflows found to date in our survey of extremely high velocity outflows (EHVO). J164653.72+243942.2 ($z_{em} \sim 3.04$) shows variable C IV$\lambda\lambda$1548,1551 absorption at speeds larger than $0.1c$, accompanied by Si IV, N V and Ly$\alpha$, and disappearing absorption at lower speeds. We perform absorption measurements using the AOD method and *SimBAL*. We find the absorption to be very broad ($\Delta v \sim 35{,}100$ km s$^{-1}$ in the first epoch and 13,000 km s$^{-1}$ in the second one) and fast ($v_{max} \sim$ -50,200 km s$^{-1}$ and -49,000 km s$^{-1}$, respectively). We measure large column densities ($\log N_H > 21.6$ [cm$^{-2}$]) and are able to place distance estimates for the EHVO ($5 \lesssim R \lesssim 28$ pc) and the lower-velocity outflow ($7 \lesssim R \lesssim 540$ pc). We estimate a mass outflow rate for the EHVO to be $\dot{M}_{out} \sim 50 - 290$ M$_\odot$ yr$^{-1}$ and a kinetic luminosity of $\log L_{KE} \sim 46.5 - 47.2$ [erg s$^{-1}$] in both epochs. The lower-velocity component has a mass outflow rate $\dot{M}_{out} \sim 10 - 790$ M$_\odot$ yr$^{-1}$ and a kinetic luminosity of $\log L_{KE} \sim 45.3 - 47.2$ [erg s$^{-1}$]. We find that J164653.72+243942.2 is not an outlier among EHVO quasars in regard to its physical properties. While its column density is lower than typical BAL values, its higher outflow velocities drive most of the mass outflow rate and kinetic luminosity. These results emphasize the crucial role of EHVOs in powering quasar feedback, and failing to account for these outflows likely leads to underestimating the feedback impact on galaxies.

## 1. INTRODUCTION

Active Galactic Nuclei (AGN) are found at the center of most massive galaxies. Among them, quasars show the largest luminosities, allowing us to study them at large redshifts and providing information about galactic evolution within our universe. Understanding the connection between the inner region and the host galaxy around has proven crucial, as evidence of a co-evolution scenario through feedback processes has been mounting in the form of the tight correlation between the masses of supermassive black holes and that of the stellar spheroids of their host galaxy ($M_{\rm bulge}$; e.g., Gebhardt et al. 2000; Merritt & Ferrarese 2001; Tremaine et al. 2002), and the need for regulating star formation in the host galaxies (e.g., Silk & Rees 1998; Di Matteo et al. 2005; Springel et al. 2005; Hopkins et al. 2006).

One promising way to connect the two regions is through outflows: gaseous material outflowing from the central engine into the galactic host. In fact, simulations have shown that cosmological feedback from AGN outflows and jets is a needed regulating mechanism (e.g., Silk & Rees 1998; Di Matteo et al. 2005; Springel et al. 2005; Hopkins et al. 2006; Ding et al. 2020). Moreover, outflows need to be understood on their own merits: they are fundamental constituents of AGN, and they

Corresponding author: Paola Rodríguez Hidalgo & Hyunseop Choi
paola@uw.edu, hyunseop.choi@umontreal.ca

provide first-hand information about the physical and chemical properties of the AGN environment and the supermassive black holes. They are detected in a substantial fraction of AGN through absorption-line signatures (e.g., broad, blue-shifted resonance lines in the UV and X-ray bands) as the gas intercepts some of the light from the central continuum source and broad emission-line region (e.g., Crenshaw et al. 1999; Reichard et al. 2003; Trump et al. 2006; Nestor et al. 2008; Rankine et al. 2020; Choi et al. 2022b, and references therein). Outflows could be ubiquitous, though, if the absorbing gas subtends a small solid angle around the background source.

For simplicity, large UV/optical surveys of quasar outflows have focused on searching for broad C IV absorption that would indicate gas outflowing at speeds less than $0.1c$; typically named Broad Absorption Line Quasars (BALQSOs). This arbitrary velocity limit was defined to avoid complications due to misidentification with Si IV or other ionic transitions blueward of the Si IV emission line. However, gas outflowing at extremely-high speeds might be the most disruptive if it is reaching the host galaxy environment, as it may provide large kinetic power. Outflows with speeds $v \sim 0.2c$, if the gas is located at similar distances and has similar physical properties, carry approximately 1-2.5 orders-of-magnitude larger kinetic power than gas outflowing at what is defined as "high" velocities ($v \sim$5,000–10,000 km s$^{-1}$) because kinetic power is proportional to $v^3$. These outflows, called extremely-high velocity outflows (EHVO; Rodríguez Hidalgo et al. 2011), might also pose the biggest challenges to theoretical models that try to explain how these outflows are launched and driven (Hamann et al. 2002; Sabra et al. 2003). Radiation pressure models (Arav et al. 1994; Murray et al. 1995; Proga et al. 2000; Ostriker et al. 2010, please see an excellent review in Crenshaw et al. 2003) take advantage of the powerful central source to accelerate line-driven winds and have proven successful in explaining different aspects of these winds such as the relation between the AGN luminosity and the terminal velocity of the outflow (Laor & Brandt 2002) and "line locking" (Turnshek 1988; Srianand et al. 2002; Hamann et al. 2011). The presence of dust mixed with the BAL gas can further boost the acceleration by radiation pressure due to the addition of scattering on dust opacity and generate outflows with velocities up to $\sim 1 - 2 \times 10^4$ km s$^{-1}$ (Ishibashi et al. 2024). However, simulations and theoretical models have faced challenges recreating the presence of detached BAL profiles with central velocities as large as $0.2c$ or greater (e.g., Proga et al. 2012; Matthews et al. 2020). Alternatively, models have also invoked magnetic forces to launch, drive, and constrain the flow (de Kool & Begelman 1995; Everett 2005; Proga & Kallman 2004), and higher terminal velocities are expected in magnetic driving due to stronger centrifugal forces (Proga 2007). EHVO must be studied if we want to understand both the central inner region and their potential effect on their galactic environment[1].

Due to the arbitrary velocity limit set in the previous systematic searches, until recently, UV/optical EHVOs had only been detected in a handful of individual quasars (Jannuzi et al. 1996; Hamann et al. 1997a; Rodríguez Hidalgo et al. 2011; Rogerson et al. 2016). Our group discovered 40 new cases (Rodríguez Hidalgo et al. 2020; hereafter, Paper I), and more recently, an additional 98 cases (Rodríguez Hidalgo in prep), by carrying systematic searches over 6760 quasars from the data release 9 (DR9Q; Pâris et al. 2012) and over 18,165 quasars from the data release 16 (DR16Q; Lyke et al. 2020) quasar catalogs of the Sloan Digital Sky Survey (SDSS) respectively. This has resulted in the first database of EHVOs and multiplied by 30 the number of known EHVO quasars.

In our DR9Q survey (Paper I), we found a very interesting case in the spectra of J164653.72+243942.2 (J1646 hereafter). J1646 shows the widest EHVO C IV absorption trough of the 40 cases found in our DR9Q survey ($\Delta v \sim$12,500 km s$^{-1}$, measured at 90% normalized flux). J1646 ($z$ = 3.040±0.002; Hewett & Wild 2010) was discovered earlier in the fifth data release (DR5) of SDSS, where it was classified as a BALQSO based on C IV absorption measured at lower speeds (Gibson et al. 2009a); this absorption disappears in the DR9 observation while the EHVO remains. J1646 is a luminous radio-quiet quasar. More information on the archival spectra and quasar properties is provided in §2.

In this paper, we analyze the absorption in the SDSS spectra of J1646 in detail using two different methods. First, we normalize the spectrum (§3.1.1) and measure the absorption by the apparent optical depth method (AOD; §3.3.1). This provides a conservative lower limit of the absorption measurements and the total column density and establishes a comparative baseline with other similar studies. Second, we utilize a state-of-the-art, novel spectral synthesis code, called *SimBAL*, that uses forward modeling to fit spectra of BALQSOs (Leighly et al. 2018), including the continuum, the emission lines (§3.1.2), as well as the absorp-tion (§3.3.2). This code is particularly well suited to fit BALQSO spectra where the absorption is too blended to be analyzed easily by other methods. As explained in Leighly et al. (2018), *SimBAL* uses grids of ionic column densities generated by the photoionization code *Cloudy* (Ferland et al. 2017) to forward-model BALQSO spectra and a Bayesian model calibration method to obtain the best-fitting values and their uncertainties. *SimBAL* uses forward modeling techniques, and a sophisticated mathematical implementation of partial covering allows accurate modeling of complex absorption features observed in BALQSO spectra (Leighly et al. 2019b). It has been used successfully in Leighly et al. (2018, 2019b), Choi et al. (2020, 2022b), and in Bischetti et al. (2024), where a *SimBAL* analysis provided robust constraints on the physical properties of LoBAL outflow in a $z \sim 6.6$ quasar. Also, Green et al. (2023) used *SimBAL* to derive physical interpretations of BAL variability seen in multi-epoch HST observations of a narrow-line Seyfert 1, WPVS 007. Finally, in Section §4 we discuss the implications of these results and comparisons to studies of absorption at lower velocities or other wavelength ranges.

---

[1] Please note that throughout this paper we will be, then, calling BALs those broad absorption troughs at a lower velocity ($<0.1c$) than EHVOs and BALQSOs the quasars that show BALs in their spectra.

**Table 1**: Date and spectral characteristics of the SDSS DR5 and BOSS DR9 data used in this work.

| DR  | MJD   | Date YYYYMMDD | Spectral Coverage Å | S/N         |
|-----|-------|---------------|---------------------|-------------|
| DR5 | 53167 | 20040611      | 3800–9200           | $20.9^b$    |
| DR9 | 55685 | 20110504      | 3600–10500          | $20.0746^a$ |

$^a$ $SNR_{1700}$ in Pâris et al. (2012)
$^b$ $SNR_{1700}$ in Gibson et al. (2009b)

## 2. DATA AND QUASAR PROPERTIES

### 2.1. Archival Spectra

The data used in this paper are archival spectra from SDSS (York et al. 2000). We discovered the special characteristics of the absorption in the spectrum of J1646 during a survey of extremely high-velocity outflows (Paper I) carried out over the SDSS Data Release 9 quasar catalog (DR9Q; Pâris et al. 2012), which was derived from the Baryon Oscillation Spectroscopic Survey (BOSS; Dawson et al. 2013) of SDSS-III (Eisenstein et al. 2011). Another previous spectrum taken ~1.70 years earlier in the quasar rest-frame was already available from SDSS DR5 (Schneider et al. 2007). Table 1 includes observation dates, wavelength coverage, and signal-to-noise ratio of the spectra used in this paper. The resolution of both spectra is 1500 at 3800 Å (2.5 Å).

Figure 1 shows both spectra (DR5 and DR9) overplotted. For easy visualization, we have smoothed both spectra with an eleven-pixel boxcar. The spectra were wavelength calibrated and sky subtracted by the SDSS pipeline. Both spectra have been dereddened for Galactic extinction using the extinction curve of Cardelli et al. (1989), assuming $R = 3.1$ and an $E(B - V)$ value of 0.0495 from Schlegel et al. (1998). The continuum is clearly higher in the early observation (DR5)[2], but the maximum absorption depth appears at a similar flux level in both observations. This is not uncommon, as we detail in Rodríguez Hidalgo et al. (in prep). The rest-frame wavelengths here and elsewhere in this paper are defined relative to the redshift $z_{em} = 3.040$ from Hewett & Wild (2010).

### 2.2. Quasar Properties

J1646 ($z = 3.040 \pm 0.002$; Hewett & Wild 2010) is a luminous quasar: SDSS provided a magnitude $g = 18.80$ (mjd = 52760) and Shen et al. (2011) calculated an absolute $i$-band magnitude, $K$-corrected to $z = 2$ ($M_i[z = 2]$) of $-28.9$ (mjd = 53167).

Rankine et al. (2020) provided improved values of physical properties of DR14 quasars by producing spectrum reconstructions covering rest frame $1260 - 3000$ Å of 144,000 quasars. These spectrum reconstructions were generated by mean field independent component analysis (ICA) using the information across the entire spectrum to inform the reconstruction. Rankine et al. (2020) provided values for J1646 of bolometric luminosity ($\log L_{\rm bol} = 47.25$ [erg s$^{-1}$]), black hole mass ($\log M_{\rm BH}/M_\odot = 9.5$), and Eddington ratio ($\log L_{\rm bol}/L_{\rm Edd} = -0.4$) using the mjd = 55685 observation. Black hole masses were calculated using the C IV emission line but accounting for the excess non-virial blue emission for quasars with large C IV blueshifts (Coatman et al. 2017). Neglecting to account for this correction results in overestimated black hole masses and, therefore, also miscalculated Eddington ratios. In fact, for J1646, the C IV blueshift is significant ($\sim 2470$ km s$^{-1}$). This seems to be typical in EHVO quasars, which show larger blueshifts than BALQSOs and non-BALQSOs in general (Rodríguez Hidalgo & Rankine 2022). The black hole mass measurements have an uncertainty of $\pm 0.5$ dex, accounting for both systematic effects and measurement errors; see Section 6.3 in Rankine et al. (2020) for more information. These values should be taken as a rough estimate due to the variability in this object; as mentioned, they were calculated using the second observation (DR9; mjd = 55685). Given that the continuum shows large variability (see Figure 1), and the flux was larger during the first observation (DR5), it is likely that the bolometric luminosity of the quasar was also larger in the first observation.

---
[2] Notice, though, that SDSS I/II fluxes have been observed to be larger than BOSS fluxes likely due to differences in the reduction process – see Margala et al. (2016) and Appendix in Shi et al. (2016).

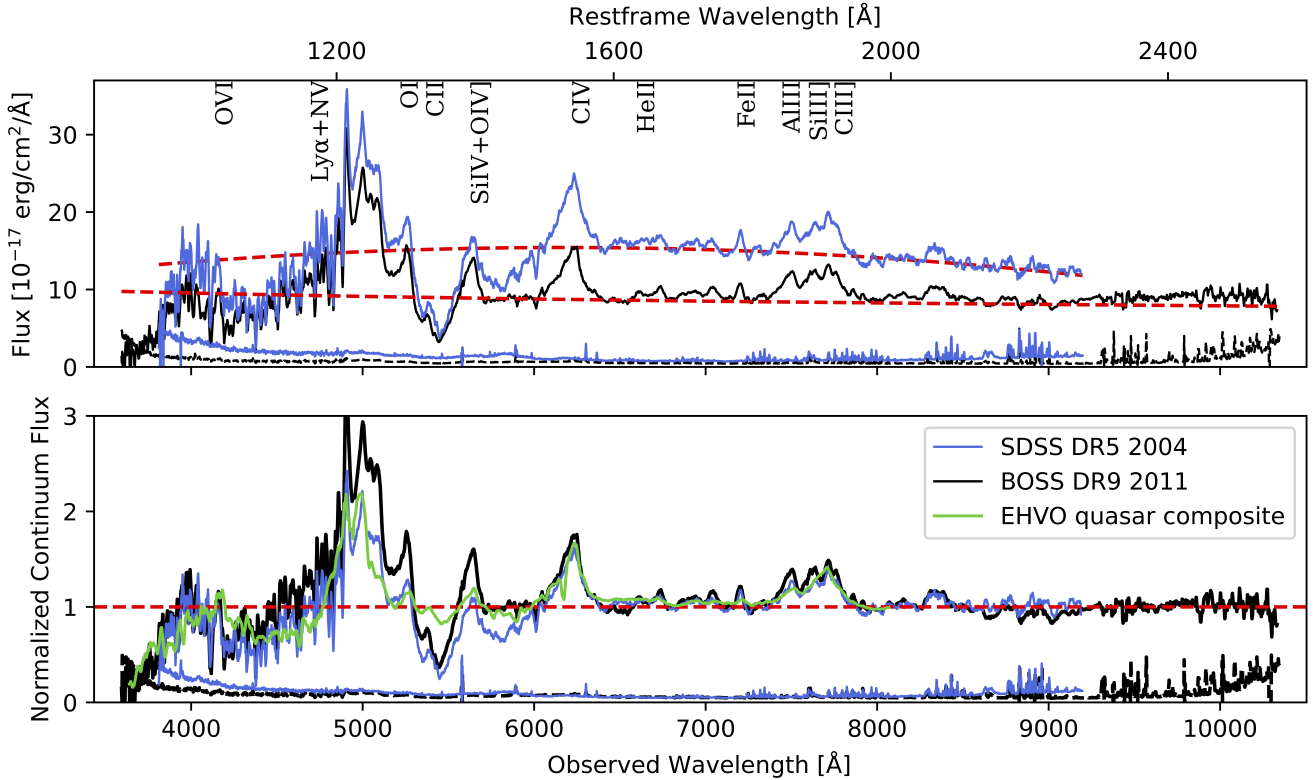

**Figure 1**: Spectra of J1646 in SDSS DR5 (blue) and BOSS DR9 (black), observed (top) and continuum normalized (bottom), with their corresponding error spectra at the bottom of each figure. All spectra have been smoothed with an eleven-pixel boxcar for easy visualization. Absorption is clearly visible in both spectra at observed-frame ∼5,400 Å with additional absorption present in the SDSS DR5 spectrum around ∼5,900 Å. Top panel: The restframe wavelength is shown at the top of the figure, and typical emission lines are indicated on top of the SDSS spectrum. The red dashed lines correspond to our continuum normalizations of the spectra in §3.1.1. Bottom: We also include a composite EHVO spectrum derived from our Paper I (green). In this case, the dashed red line is included to indicate the normalized continuum level and help guide the eye.

J1646 is a radio-quiet quasar as it has no identified match to a Faint Images of the Radio Sky at Twenty-Centimeters source included in Pâris et al. 2012.

## 3. ANALYSIS & RESULTS

### 3.1. *Normalization of the quasar spectra*

To perform measurements of the absorption, first we normalized both spectra. Quasars tend to show negative-sloped spectra in the UV/optical region, and composite quasar spectra are well approximated by a power-law in this region, even in the far UV (e.g., Vanden Berk et al. 2001; Telfer et al. 2002; Tilton et al. 2016). In Paper I, we performed a systematic normalization of the 6760 quasar spectra in the sample, focusing on the wavelength region between the Ly$\alpha$+N v and C iv emission lines, in order to search for and measure EHVO absorption in that wavelength range. For the work in this paper, we have refined the normalization of the J1646 spectra to include the Ly$\alpha$ forest region, and a more tailored fitting of the region redward of the C iv emission line.

Throughout this paper, we show two different procedures, performed independently, to characterize the absorption in the J1646 spectra: (1) the Apparent Optical Method (AOD), for which only the underlying continuum is fitted; this methodology provides a solid lower limit for the outflow measurements, and (2) a best estimate one, using *SimBAL* (described in §1), where the continuum, emission and absorption are fitted simulta-neously. In subsequent sections we describe how we used both methods to analyze the continuum and emission lines in the spectra (§3.1.1 and §3.1.2), identify (§3.2), and measure the absorption (§3.3.1 and §3.3.2).

### 3.1.1. *Apparent Optical Method approach: continuum fitting and emission line analysis*

As can be seen in Figure 1, the continua between both epochs differ not only in flux level but also in slope. Therefore, we followed different approaches to fit the continuum for each spectrum.

For the BOSS DR9 2011 spectrum, we used a simple power law (similar to what is shown in Fig. 6 of Vanden Berk et al. 2001) anchored at four points in the quasar spectra. All anchor points were selected away from emission and absorption features. The power law seems to match the continuum correctly throughout the spectrum.

For the SDSS DR5 2004 spectrum, a single power law did not produce an adequate fit as the continuum seems to curve down at shorter wavelengths. This is not rare: the composite spectrum by Telfer et al. (2002), which was made from a sample of 332 HST/FOS quasar spectra, required a broken power law to fit the continuum blueward of the Ly$\alpha$+N V emission line. Then, a simple power law would have overestimated the absorption in this region.

The normalization of the Ly$\alpha$ forest is complicated due to (1) the combination of a myriad of hydrogen absorption lines and the presence of emission lines such as O VI $\lambda\lambda 1032,1038$ and (2) in our case, being a narrow wavelength region located at the edge of the SDSS/BOSS spectrum where the sensitivity and SNR are reduced. Thus, for both spectra, we tried to extrapolate a reasonable fit from the wavelength region redward of the Ly$\alpha$+N V emission line into the Ly$\alpha$ forest without selecting any point to anchor the fitting in this complex region. For the DR9 spectrum, the simple power law produces a reasonable continuum in this region. For the DR5 spectrum, a 2nd-order polynomial function mimics the broken power-law and allows us to extrapolate the fitting blueward of the Ly$\alpha$+N V emission line into the Ly$\alpha$ forest. All of the anchoring points were selected redward of the C IV emission line since the region between the Ly$\alpha$+N V and C IV emission lines ($\sim$1200 - 1600 Å in the rest-frame) contains an entangled combination of emission lines and broad absorption.

Figure 1 also shows the normalized spectra that were obtained by dividing the original spectra by each continuum fitting. While both normalizations were carried out independently and through different methods, as explained above, many of the emission lines overlap, including the O VI emission line. We include as well a composite (green) derived from our survey of EHVO quasars in Paper I. This weighted-mean composite is created by combining the 16 cases with 2.5< $z_{em}$ <3.25 (approx. the middle third of our DR9 sample), which we resampled interpolating to a common wavelength grid prior to averaging. We selected these quasars for the composite because our survey of EHVO DR9 quasars found a large range of redshifts, and the Ly$\alpha$ emission line shows very different contaminations with Ly$\alpha$ forest lines depending on quasar redshift. We do not mask the absorption, and thus, it appears averaged in the composite. This approach is conservative, and we might be underestimating the continuum levels of both spectra, especially the SDSS DR5 spectrum, if additional absorption is present. However, it is very unlikely that we have overestimated the continuum levels: in Telfer et al. (2002), the elbow of the broken power-law occurs at the Ly$\alpha$+N V emission line, not redward of it. Absorption measurements using this method will be, thus, lower limits.

While we do not fit the emission lines within this approach, we took advantage of having two epochs of spectra and noticed that the emission lines are very similar between epochs. Figure 1 shows also the emission lines in the 6000-8000 Å region (C IV, He II, Fe II, Al III, C III], labeled at the top of the figure) resemble each other quite well and they seem to differ only by a scaling factor.

Even in the region where absorption is clearly present, emission lines appear to retain their shape and to be displaced downwards. In the DR9 BOSS spectrum, C II could be part of the absorption profile. In the DR5 SDSS spectrum, O I, C II and Si IV+O IV] emission lines may be embedded within the absorption profile; this is very clear in the case of the Si IV+O IV] emission line, which appears to be surrounded by absorption (see Fig. 1 $\sim$1400 Å in rest-frame, $\sim$5700 Å in observed wavelength). Without access to a different epoch, it could have been interpreted as if the emission line had been completely absorbed.

Using the similarities between emission lines in the two epochs, which correspond to a wide range of physical conditions, we took a comparative approach. We assumed that the absorption in the DR9 spectrum is restricted to absorbing the continuum and that the emission lines surrounding absorption features (such as O I and Si IV+O IV]) are not affected by it. Figure 2 shows the method we used. We created a template of emission lines, which we call the scaled DR9 template, by interpolating the BOSS spectrum to the wavelength grid of the SDSS spectrum and using the region between rest-frame 1500 and 1950 Å (between the C IV and Al III+C III] emission lines) to determine the scaling factor (Figure 2

- top). Using the normalized spectra, the scaling factor was the value that minimized the median of the values for *[(BOSS DR9 - 1) × (scaling value) + 1] - SDSS DR5*, in this region. Figure 2 (top) shows the scaled BOSS DR9 spectrum (blue) overplotted together with the SDSS DR5 spectrum. Figure 2 (middle) shows how close the emission lines of the scaled DR9 match the DR5 spectra. Finally, we divided the SDSS DR5 spectrum by the scaled BOSS DR9 spectrum.

Figure 2 (bottom) shows the result of this process. The absorption in magenta represents the excess absorption in the DR5 spectrum relative to the absorption in the scaled version of DR9; in other words, it is partly the absorption that disappears between epochs. It shows that there is continuous absorption starting at ∼1480 Å up to ∼1050 Å, together with the previously distinct EHVO. The nature of the absorption is investigated in the following sections.

To ensure this very-wide absorption trough was not some sort of glitch in the SDSS DR5 spectra of this object, we inspected the individual spectral exposures from the SDSS red-arm and blue-arm spectrographs. Each of the three individual exposures from each spectrograph is consistent with the others within the noise. In the region in observed wavelengths from 5825 Å to 6150 Å where the spectra from the two arms overlap, the average blue-arm spectrum exhibits a $\simeq 15\%$ lower flux level than the average red-arm spectrum, at a significance of $3.4\sigma$ assuming Gaussian statistical errors only. However, we do not believe this is strong evidence against the reality of the apparent very wide absorption trough in the DR5 spectrum. First, we do not see any blue-arm flux offsets on the two neighboring spectra on the plate. Second, multiplying the normalized DR5 spectrum by a factor of 1.15 before comparing it to the scaled DR9 spectrum does not eliminate the putative absorption trough, particularly at wavelengths just longward of Si IV. Third, the apparent very wide absorption trough absorbs an approximately fixed fraction of the normalized quasar *continuum*, but any plausible error in flux scaling will result in a difference spectrum feature that is a fraction of the observed *spectrum*, not just the continuum. Furthermore, the very wide absorption trough does not span the full wavelength range of the spectrum from the blue-arm spectrograph. Although a negative offset to the blue-arm spectrum, which happened to be constant in $F_\lambda$ could match the difference spectrum given the relatively flat-in-$F_\lambda$ spectrum of the quasar, such an offset would spoil the match between the normalized DR5 and DR9 spectra at the shortest wavelengths common to both spectra.

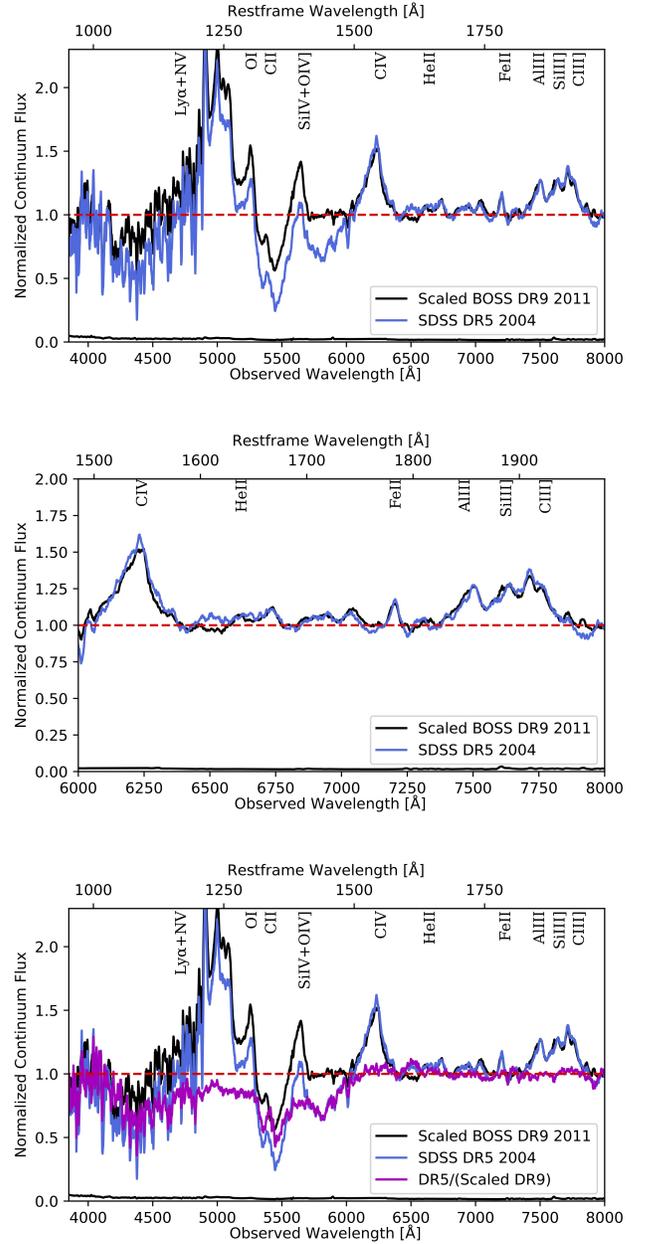

**Figure 2**: Top & Middle – Scaled BOSS DR9 spectrum (black) overplotted over the SDSS DR5 spectrum (blue), showing the whole spectra (top) and only the region used to find the scaling factor (middle). Bottom – Original DR5 spectrum (blue) and masked and scaled BOSS DR9 spectrum (black) overplotted together with the divided spectrum of the two (magenta). All spectra have been smoothed by an 11-pixel boxcar as well. Error spectra are shown below each spectrum.

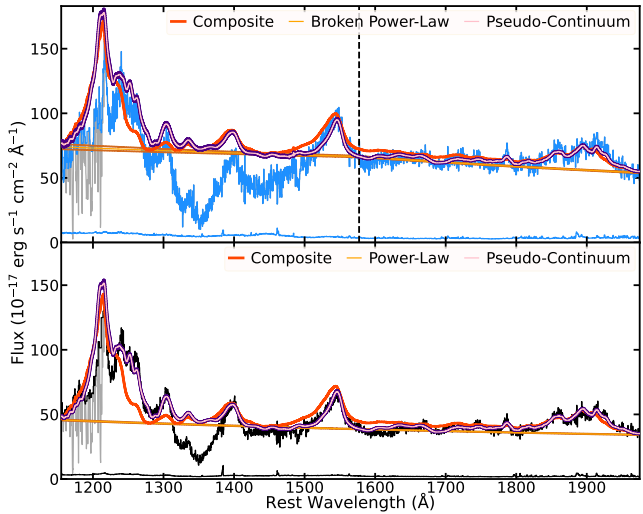

**Figure 3**: The pseudo-continuum model components extracted from the best-fitting *SimBAL* models for SDSS DR5 (top) and BOSS DR9 (bottom). Our emission line models generated from the SPCA eigenvectors show realistic rest-UV quasar emission lines with line ratios and strengths comparable to what are found in a composite spectrum. The grey points at $\lambda \lesssim 1216$ Å show the data points affected by the Lyman $\alpha$ forest features identified by our iterative method, and they were ignored for *SimBAL* fitting (see Appendix A for details). The emission composite from Temple et al. (2021) is plotted in red. The vertical line in the top panel shows the location of the slope break point ($\lambda \sim 1690$ Å) for the broken power-law model used in SDSS DR5.

Absorption measurements in section § 3.3.1 are performed over the normalized DR5 and DR9 spectra. The findings in this section show that we are likely underestimating the amount of absorption, but it sets a firm, conservative, lower limit for it, which is our overall goal with this method.

### 3.1.2. *Best estimate: continuum + emission lines fitting using SimBAL*

*SimBAL*, when it was first developed, was used to model the absorption lines alone (see Leighly et al. 2018). Instead, in this paper, as in Choi et al. (2020, 2022b), we modeled both the pseudo-continuum (continuum + emission features) and absorption lines simultaneously. Choi et al. (2022b) introduced the use of spectral principal component analysis (SPCA) eigenvectors for the emission line modeling for rest-UV spectra within *SimBAL*, which we also included in this work. SPCA pseudo-continuum modeling can reproduce realistic emission line features for a given wavelength range with fewer model parameters than individual line-fitting procedures, which require multiple parameters per emission line in the model. While *SimBAL* carries out the fitting of the pseudo-continuum and the absorption simultaneously, to compare with our AOD method, we describe the former in this section and the latter in section §3.3.2.

We used a power-law model for the continuum emission for the BOSS spectrum and a broken power-law model for the SDSS spectrum to model the observed spectral shape. The slope break point for the broken power-law model was constrained $\lambda \sim 1690$ Å from the *SimBAL* fit. To fit the emission lines, we used three sets of eigenvectors because we currently do not have a single set that spans the whole SDSS/BOSS spectral region. We used one set for emission lines between 1030 and 1290 Å and another for emission lines between 1290 and 1700 Å. The first set was constructed by us using the sample of 78 $z \sim 3$ quasars discussed in Pâris et al. (2011) and the second set was made from a sample of $\sim 100$ quasars that show strong blueshift in C IV emission lines, similar to what we observe in J1646 (Hazlett et al. 2019). Redward of 1700 Å, we used the eigenvectors described and used in Choi et al. (2022b), which were built from a set of 2626 SDSS non-BAL quasars. Each set of SPCA eigenvectors has six parameters: four coefficient parameters that yield the shape and the line ratios of the emission lines and two additional parameters, the convolutional width parameter and the amplitude parameter, that control the overall strengths and the widths of the emission lines. The width parameter is not required for the standard eigenvector reconstruction. Nevertheless, we included the parameter in our model to provide it with an additional method to reproduce a broad range of emission line widths observed in quasar spectra.

Figure 3 shows the pseudo-continuum models extracted from the best-fitting *SimBAL* models for SDSS DR5 and BOSS DR9. We compared our model with a composite template to examine whether our emission line model from spectral PCA eigenvectors has produced reasonable emission line ratios and shapes that are comparable to those observed in real data. Temple et al. (2021) provides two emission-line templates ("high-blueshift" and "high-equivalent width") from their quasar Spectral Energy Distribution (SED) model. We chose the "high-blueshift" template, which showed a good match with the data and our pseudo-continuum models around C IV and C III] emission line regions with similar line flux ratios between major emission lines (e.g., Ly$\alpha$, Si IV, C IV). The "high-equivalent width" template was disfavored as it showed a high C IV/Si IV flux ratio that we do not see in our target (see Figure 4 in Temple et al. 2021). Furthermore, since EHVOs exhibit larger C IV blueshifts in emission (Rodríguez Hidalgo & Rankine 2022), the high-blueshift template from Temple et al. (2021) is the most suitable for our target among the available composites (e.g., Vanden Berk et al. 2001; Francis et al. 1991). We performed an additional continuum emission subtraction between $\sim 1440$ Å and $\sim 2000$ Å because the original template showed unusually strong Fe II emission features throughout that bandpass, and we suspected that it was contamination from inaccurate continuum subtraction. The modified emission line template was then scaled to match the strengths of the emission lines that are not affected by BAL, such as C IV and the emission lines around C III], and we added the scaled template to our continuum models.

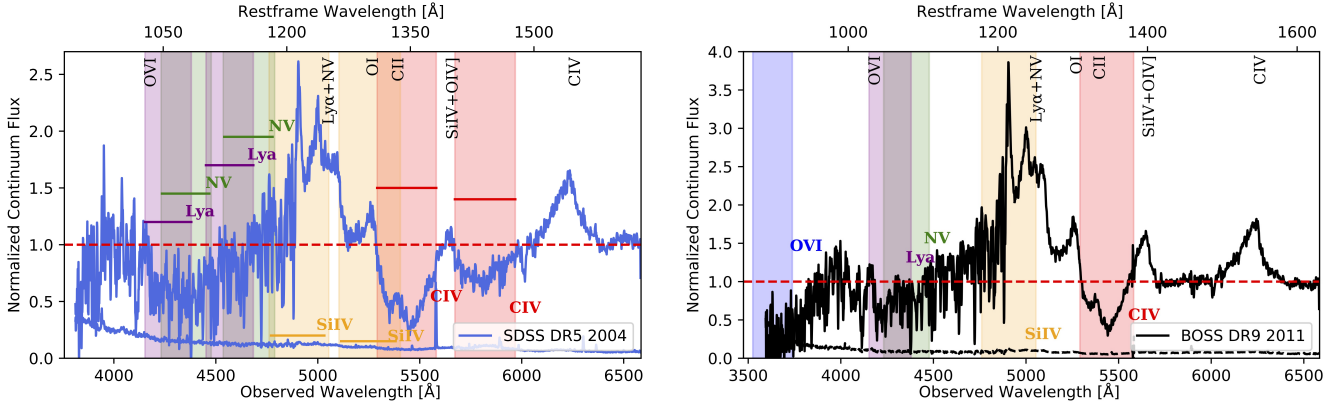

**Figure 4**: Spectra of J1646 in SDSS DR5 (left) and BOSS DR9 (right), including the location of potential ions with different colors. EHVO C IV absorption is indicated by a red-shaded region; the upper and lower velocity limits are determined by the wavelengths where the absorption crosses the 90% level of the normalized flux. The DR5 spectrum also includes C IV absorption at lower velocities. The blue, purple, green, and orange shaded regions have the same lower and upper velocities as the red region but shifted to the corresponding O VI, Ly$\alpha$, N V and Si IV wavelengths, respectively; $v_{max}$ and $v_{min}$ are included in Table 2. Although the Ly$\alpha$ forest is a complex region and lies at the edge of the spectrum, the width of the absorption and the changes between the two spectra suggest that EHVO absorption is present in both cases. O VI absorption cannot be confirmed in either spectrum because it lies at the spectral edge in DR9 and is not covered in DR5. Some emission lines present in the spectrum are labeled at the top of each figure. For clarity, in the case of the DR5 spectrum (left) we also include colored horizontal lines to mark the limits of each shaded region.

We found that our SPCA reconstructed pseudo-continuum models match very closely with the ones made with the emission-line template. Although our object (and the models) shows slightly stronger low-ionization emission lines (e.g., Si II), major emission lines show a good match. We note that the template shows a slightly higher C IV blueshift than what is observed in our pseudo-continuum models and the data. This verifies that our SPCA reconstruction method indeed produces realistic emission line models and also shows that J1646 has emission line properties similar to those of quasars with highly blueshifted emission lines. For the *SimBAL* analysis, we employ the SPCA reconstruction method to model the pseudo-continuum. The template was used solely for comparative purposes in Figure 3.

In EHVO quasars, Si IV absorption lines are often located on top of Ly$\alpha$+N V emission lines, making it difficult to estimate the true strengths of the emission complex and the amounts of Si IV opacity from the BALs. Choi et al. (2022b) found that in some BAL quasars, the emission lines were not absorbed by the BAL. The same phenomenon was also observed in WPVS 007 (Green et al. 2023). We modeled J1646 with and without emission line absorption and found that a spectral model with no emission line absorption produced a more self-consistent fit. The models with emission line absorption predicted strong Ly$\alpha$+N V emission features that are not found in EHVO quasars (see composite described above). This was because the models predicted a moderate amount of Si IV opacity near the Ly$\alpha$+N V region and, in order to match the flux levels observed in that region, the models produced strong emission lines to compensate for the absorption from Si IV BAL. In contrast, the models with no emission line absorption from the BAL produced pseudo-continuum models that resemble typical emission features seen in EHVO quasars. While this modification to the models resulted in much weaker Lyα+N v emission lines in the pseudo-continuum models than the initial model we tried, the overall BAL physical properties did not show significant differences. We note again that the continuum and line emission fitting was done simultaneously with the BAL absorption modeling.

We did not find evidence for significant emission line variability between SDSS and BOSS (see Figure 1). Therefore, we simultaneously fit both the SDSS and the BOSS spectra, constraining the values of eigenvector coefficient parameters and the width parameters of the emission lines to be the same for both SDSS and BOSS spectra models. The continuum model parameters and the emission line amplitude parameters were allowed to vary between the models for SDSS and BOSS, such that the equivalent widths of the emission lines were allowed to be different for the two models while keeping the shapes of the line profiles identical.

### 3.2. Identification & Nature of the Absorption

Figures 1 and 3 show absorption present below the continuum at rest-frame wavelengths ∼1,360 Å in both DR9 and DR5 spectra. The absorption present in the BOSS DR9 2011 spectrum was classified as an EHVO C IV in Paper I.

In the DR5 SDSS 2004 spectrum, an absorption feature at ∼1,440 Å is also clearly present. Due to the wavelength range, the most likely interpretation is as C IV outflowing at shorter velocities. Thus, this quasar would be a C IV BALQSO in the typical definition (Weymann et al. 1991) in DR5, and so it was classified (Gibson et al. 2009a[3]). This trough disappears within 1.7 yrs in the quasar rest-frame, as it does not seem to be present in the DR9 spectrum.

#### 3.2.1. AOD method: Visual Identification of Other Ionic Transitions

Figure 4 includes the normalized spectra, where just the continuum has been normalized, for SDSS DR5 (left) and BOSS DR9 (right), where we have labeled the potential location of other ionic transitions. The limits of each colored region are set by the C IV absorption detection below 90% of the flux level in the normalized spectrum (red; see §3.3), assuming that the other ions are found at the same velocities. Due to the large absorption widths, no doublets are present to help confirm the nature of any of the absorption troughs, but C IV HiBALs are typically accompanied by N v, Si IV, Lyα, and/or O VI in the same outflow (in Fig. 4, green, or-ange, purple, and blue shaded regions, respectively), so we searched for these and other transitions at the expected wavelengths assuming they would be at similar velocities.

---
[3] Gibson et al. (2009a) incorrectly classified the EHVO C IV as "Si IV" as it is typically the case in previous surveys.

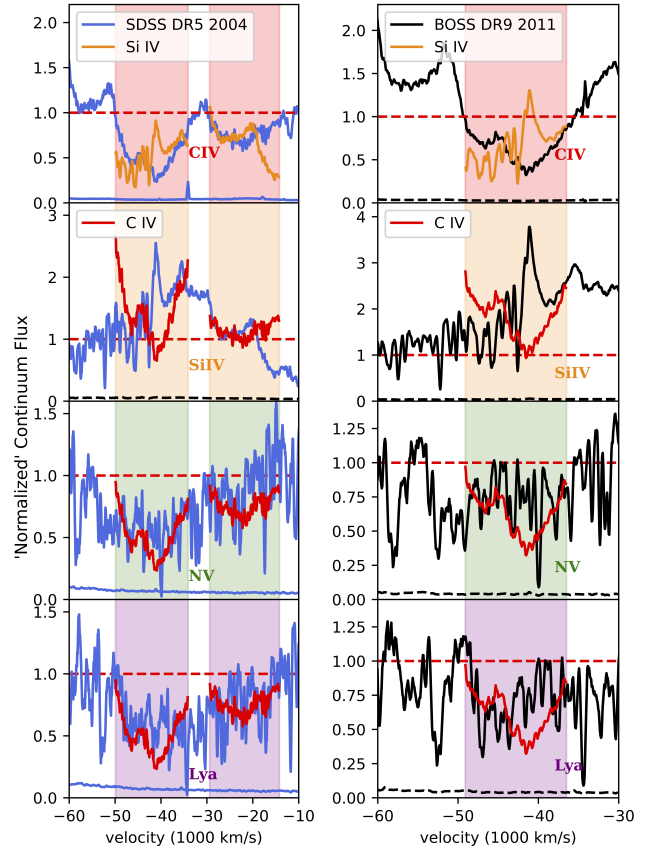

**Figure 5**: Normalized spectra (DR5 left – blue, DR9 right – black) shown in velocity space at the location of C IV (red bars, top panel in each group), and the potential detections of Si IV (orange bars, second panel), N v (green bars, third panel) and Lyα (purple bars, bottom panel). The shaded regions are defined based on the C IV absorption and transported to the potential location of the other ions. The superimposed red and orange solid lines represent the C IV and Si IV absorption profiles, respectively. The detection of EHVO Si IV lies on top of the Lyα+N v emission, but the similarity between the profiles, especially at the lower EHVO velocities in both SDSS and BOSS spectra, suggests that some Si IV absorption is present. Lyα and N v absorption overlap in spectral wavelengths and appear to be both present, but Lyα seems to be stronger, especially at the higher-velocity tail of the EHVO absorption in both epochs.

Figure 5 shows the absorption profiles of these ions in velocity space. The left four panels show the absorption in the DR5 spectrum, and the right four in the DR9 spectrum. The top panels show the C IV absorption (red shade) with the over-plotted Si IV profile (orange line) at the same outflow speed and the Si IV absorption (orange shade) with the over-plotted C IV (red line). The bottom panels show the potential absorption of N V (green shade) and Ly$\alpha$ (purple shade), all with the C IV profile over-plotted (red line). In the case of doublets, the velocity scale corresponds to the short wavelength components of the doublets relative to the emission redshift. The shaded regions are defined based on the C IV absorption and transported to the potential location of the other ions.

In Figure 4, it is difficult to assess whether the absorption at the wavelengths $\sim$1,100 Å is due to N V or Ly$\alpha$, or a combination of them, as they overlap at these absorption widths. Figure 5 (bottom left plots) shows that in the DR5 spectrum, both ionic transitions are likely to show some EHVO absorption. In both SDSS and BOSS, the profiles of Ly$\alpha$ and C IV follow the same shape (with the exception of Ly$\alpha$ forest lines) at the highest speeds of the EHVO absorber (between $\sim$-50,000 km s$^{-1}$ and $\sim$-43,000 km s$^{-1}$), and N V absorption is also likely present at $\sim$40,000 km s$^{-1}$ in SDSS. The absorption at a lower velocity ($\sim$20,000 km s$^{-1}$) does not present a good match between C IV and either of these two species, but the absorption is also weaker overall. In the DR9 spectrum (see Fig. 5 bottom right plots), EHVO Ly$\alpha$ is likely to be stronger than N V but weaker than C IV, and only the higher-velocity part of the EHVO profile appears to show potential absorption of these two species (between $\sim$-49,000 km s$^{-1}$ and $\sim$-47,000) km s$^{-1}$. In summary, the bottom plots in Fig. 5 suggest that both Ly$\alpha$ and N V might be present, that their absorption profiles differ from that of C IV in that they have relatively more absorption at the highest velocities, and that Ly$\alpha$ absorption might be stronger than N V, especially at the higher-velocity tail of the EHVO absorption.

Based on our previous studies, we find it surprising that Ly$\alpha$ absorption is stronger than N V. In Rodríguez Hidalgo et al. (2011), we presented the analysis of other accompanying ions in the spectrum of an EHVO quasar, where N V and O VI appeared to be present, but Ly$\alpha$, if present, was not strong and blended with Ly$\alpha$ forest lines. Other EHVO quasars in Paper I also seemed to show strong N V absorption, while Ly$\alpha$ absorption was much rarer than N V. Analysis of the composite shown in Fig. 1 (bottom) shows how the average absorption in this region ($\sim$ 1,100 Å in the restframe) favors the interpretation of N V and that the N V absorption is in average stronger than C IV absorption. An alternative interpretation could be that the C IV absorption is broader as it appears currently framed by two emission lines, O I and Si IV+O VI], that could be suffering some additional absorption we have not accounted for, as we discussed in §3.1.1. If this is the case in J1646, it might be indicating either that the N V absorption is wider than the C IV absorption or that the C IV absorption is wider as well and partly present on top of the Ly$\alpha$+N V+Si II emission line complex. Indeed, comparisons between the DR5 and DR9 spectra show that emission lines around the EHVO C IV absorption appear to be absorbed in the DR5 (Fig. 1), while emission lines such as C IV and Al III+C III] are almost overlapping between the two spectra. Using the scaled DR9 template, the SDSS DR5 spectrum (see Fig. 2) showed that absorption may be present over a larger range of wavelengths, and it is even more complicated to determine its nature. However in J1646, we prefer the original interpretation of Ly$\alpha$ being stronger than the N V absorption because using $SimBAL$ we experimented with a larger starting velocity of $\sim$-64,800 /kms for the C IV absorption, and it converged to the best estimate where the C IV maximum outflow velocity is $\sim$50,000 km s$^{-1}$ as in Fig. 5 (see next section 3.2.2). Besides "atypical" Ly$\alpha$ absorption, this quasar was reported in Paper I to be one of the three, out of 40 quasars, to show "atypically strong" He II $\lambda$1640.42 emission (please see their section 5.3). More studies on EHVO quasars are necessary to determine what is "typical" for these quasars.

Potential Si IV absorption lies on top of the Ly$\alpha$+N V emission line complex (see Fig. 4), which also makes its identification more difficult. In Figure 5 (left plots), we can observe that the shape of the profiles of C IV and Si IV resemble each other at velocities $\sim -38,000$ to $-35,000$ km s$^{-1}$ for both DR5 and DR9 spectra, as well as at the higher-velocity limit for high-velocity absorption in DR5 ($\sim -25,000$ km s$^{-1}$). This suggests that some Si IV absorption is likely present.

Potential O VI absorption appears at the edge of the BOSS DR9 spectrum, and it is not covered in the SDSS DR5 one, so it cannot be studied and it is not included in Fig. 5.

### 3.2.2. *Best estimate: spectral modeling of the absorption with SimBAL*

We simultaneously fit both DR5 and DR9 using $SimBAL$ and compare the constrained physical properties of the outflow gas seen in both epochs. $SimBAL$ uses six parameters to model the absorption features: ionization parameter $\log U$, density $\log n$ (cm$^{-3}$), column density parameter $\log N_H - \log U$ [cm$^{-2}$], outflow velocity $v$ (km s$^{-1}$), velocity width $\Delta v$ (km s$^{-1}$), and a dimensionless covering fraction parameter $\log a$ (higher value corresponds to a lower covering fraction). Unlike $C_f$, which is used for homogeneous partial covering, the $\log a$ parameter models the inhomogeneous partial covering of the pseudo-continuum emission by using a power-law opacity profile of the BAL gas (e.g., Arav et al. 2005; Sabra & Hamann 2005). We used the "tophat" setting in *SimBAL*, which models the broad trough with rectangular bins of equal velocity width to span the BAL. This approach enables detailed extraction of outflow physical parameters as a function of velocity. We refer to Leighly et al. (2018, 2019b) for a detailed discussion on *SimBAL* modeling methods, including the physical interpretations of the power-law partial covering.

Following the method pioneered in Green et al. (2023), we simultaneously fit both epochs, tying all absorption parameters between epochs except one to see if and which single parameter explains the variability observed between DR5 and DR9. Figure 6 shows our best-fitting *SimBAL* models for the two epochs, in which only the covering fraction parameter ($\log a$) was allowed to vary between the two epochs while other physical parameters, such as $\log U$, $\log N_H - \log U$, were constrained to be identical between the models for the two epochs. We tested other scenarios where we allowed ionization parameter ($\log U$) and column density ($\log N_H - \log U$) to vary. Although all three models produced comparably acceptable fits, the model incorporating varying covering fraction parameters yielded the best fit, particularly in reproducing the deep C IV EHVO profile. Green et al. (2023) tested various scenarios to explain the variability observed in WPVS 007, a low-redshift Seyfert 1 galaxy with UV-BAL variability, and found, consistent with our results, that changes in the covering fraction ($\log a$) were the primary driver of the observed variability. Similarly, they found that models assuming variability driven solely by changes in either the ionization parameter or the column density failed to reproduce the deep P V and Si IV BAL features observed in their object. Thus, we took the simplest variability model, in which only the covering fraction parameters vary between the epochs, as it produced robust fits to the data and the simplest explanations for the variability with few assumptions. We discuss the implications in § 4.3.3.

The best-fitting models for J1646 have the following prescription:

$$f_{model}(\lambda) = f_{continuum}(\lambda) \times I_{BAL}(\lambda) + f_{line\ emission}(\lambda),$$

where the power-law (or broken power-law for SDSS) continuum emission is absorbed by the outflow gas ($I_{BAL}$) and the emission lines are not absorbed by the outflow gas (for our reasoning, please see section §3.1.2). As described in Leighly et al. (2018), the *SimBAL* analysis begins with an initial manual fitting of the spectrum by adjusting the fit parameters to get the *SimBAL* model to roughly match the data. We began the fitting process with the visual inspection of the spectra to estimate the outflow velocity of the BAL in order to place the bins for modeling the absorption. Based on Figures 4 and 5, we determined that the C IV trough starts from near the O I/S II emission lines and ends before the Si IV emission line for DR9 and before the C IV emission line for DR5. Due to their different widths, we used a 19-bin tophat model to fit the troughs in the DR5 J1646 data and an 8-bin model for DR9 to fit the troughs in DR9. We kept the bin width and the velocity of the highest-velocity bin the same between DR5 and DR9 spectral models so that we could directly compare the differences in the column density and covering fraction parameters between the two epochs as a function of velocity. The number of bins (or bin width) used in the tophat model does not affect the robustness of the *SimBAL* model fits (Leighly et al. 2018).

For this object, the tophat bins were constrained to have the same $\log U$ and $\log n$ across the velocities while allowing $\log N_H - \log U$ and $\log a$ for each bin to freely vary to model the BAL features. However, we did not include density as a fit parameter because we do not have any density-sensitive diagnostic absorption lines (e.g., Lucy et al. 2014) that can be used to constrain the density from the model fitting; instead we fixed the density parameter at $\log n = 6$ (cm$^{-3}$). As mentioned, both epochs were fit simultaneously with all the absorption parameters shared between the spectral models for DR5 and DR9, except for the covering fraction parameter. However, the BAL feature observed in DR5 shows a much larger velocity width than in DR9, extending to lower velocities (§ 3.2.1). We assigned a separate $\log U$ parameter for the modeling of the lower velocity portion of the BAL trough in DR5.

We used *SimBAL* to fit both the pseudo-continuum (emission lines + continuum) and the absorption lines together (§ 3.1.2). We ran *SimBAL* to obtain the converged MCMC chain, and using this chain, we generated best-fitting models for both epochs and extracted posterior distributions of the parameters with which we obtained the physical properties of the outflows. The principal absorption lines found by the best-fitting *SimBAL* models are C IV, Si IV, N V, and Ly$\alpha$. Visual inspection did not reveal low-ionization transitions in either spectrum (such as Mg II and Al III), but the best-fitting models found evidence for weak Al III, indicative of high column density wind (e.g., Bischetti et al. 2024).

The best-fitting model for DR5 shows a very broad absorption feature stretched from $\sim 1500$ Å to $\sim 1180$ Å (third panel from the top in Figure 6). The feature consists of blended C IV and Si IV that overlap between

12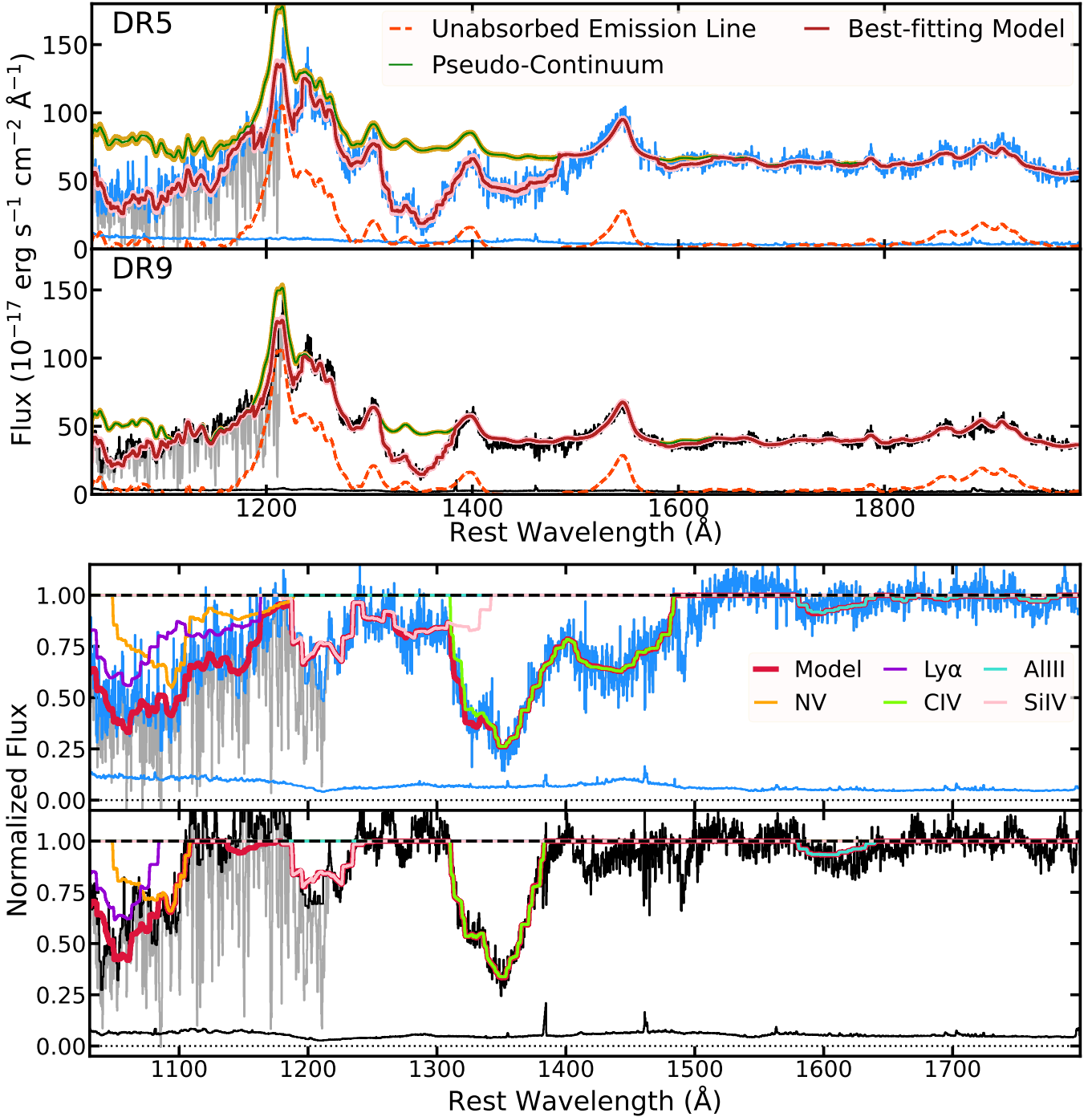

**Figure 6**: Top two panels: the best-fitting *SimBAL* models and uncertainties (red and pink, respectively) for SDSS DR5 (blue) and BOSS DR9 (black), plotted over the rest-frame wavelength range of $1050-2000$ Å used in the modeling. Bottom two panels: the pseudo-continuum normalized models with the line identifications, shown over a zoomed-in range of $1050-1800$ Å. Grey lines show the spectrum data that includes both the outflow BAL troughs and the Lyman forest absorption lines, and the black lines show the data we used for the *SimBAL* model fitting, where the non-BAL absorption lines have been flagged and ignored. The unabsorbed emission line features (gold) reveal a large amount of unabsorbed flux in the Ly$\alpha$+N v emission line region (§ 3.1.2). The BAL trough extends to a lower velocity with greater width in the SDSS data. The model for BOSS shows a little opacity from low-ionization ions such as Al iii, indicating a high column density outflow.

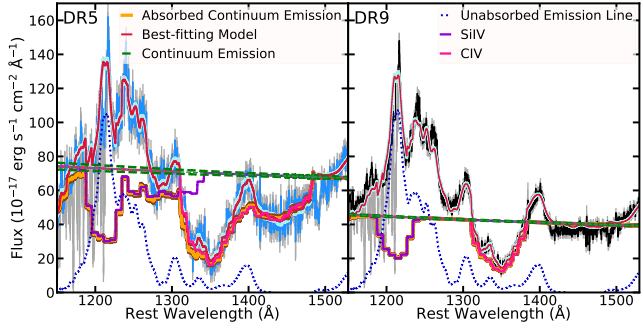

**Figure 7**: Si IV absorption identified by best-fitting *SimBAL* models in DR5 (left) and DR9 (right). A considerable amount of absorption from Si IV is seen in the continuum emission-only models, appearing much deeper than observed in Figure 6 due to strong Ly$\alpha$+N V emission filling the bottom of the troughs (§ 3.2.2). The opacity profile of Si IV largely resembles that of C IV.

$\sim 1300$ and $\sim 1350$ Å. We experimented with a modified *SimBAL* model for DR5 with extra tophat bins at the highest velocity end in order to test the possibility that the absorption corresponds completely to an ultra-wide trough of C IV. The converged model from this experiment showed no discernible differences from our best-fitting model; in other words, even when inputting a larger velocity of C IV to the test model, the parameters rearranged themselves to create Si IV opacity and only show significant opacity C IV in between $\sim 1300$ Å and $\sim 1500$ Å, identical to what we found in the best-fitting model. Therefore, we eliminate the possibility that the absorption is wider than presented in the best-fitting model.

In order to investigate how much Si IV opacity is hidden near the Ly$\alpha$+N V emission line, we separated the emission lines and continuum emission from the best-fitting models (Figure 7). The model decomposition clearly shows deep Si IV troughs in the accretion disk continuum emission. In contrast, the best-fitting *SimBAL* models in Figure 6 only show a moderate amount of apparent Si IV opacity from the main EHVO. That is because the bottoms of the Si IV troughs have been filled in by the flux from the unabsorbed Ly$\alpha$+N V emission line, making the apparent depths of Si IV troughs shallower. This type of behavior is also seen in objects with lower velocity outflows, where Ly$\alpha$+N V BAL is filled in by the emission lines (e.g., Leighly et al. 2019a; Green et al. 2023).

### 3.3. *Measurements of Absorption*

We also followed the two main approaches to characterizing and measuring the absorption. The AOD method serves as a lower limit and comparison to a more traditional method for the results obtained through *SimBAL*. One of the main differences is that the constraints from the AOD method are obtained from the C IV BAL features alone, whereas *SimBAL* spectral modeling takes into account the entire wavelength range.

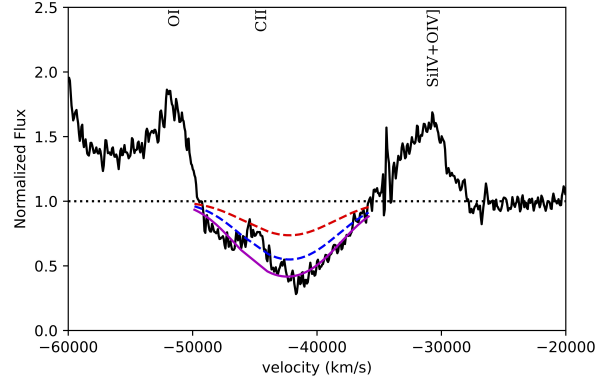

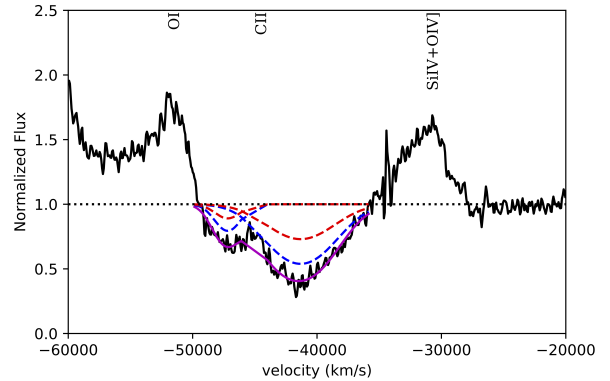

**Figure 8**: Examples of fitting results for the DR9 spectrum using our iterative approach (methods 2 and 3 in Table 3). Method 2 (top): we use one pair of Gaussians for the C IV doublet. Method 3 (bottom): we use two pairs of Gaussians. In both cases, blue corresponds to C IV 1548Å and red to C IV 1550Å, and the combined fitting is indicated by the solid purple line. The region where C II 1335Å emission seems to be present ($-46{,}000 < v < -44{,}000$ km s$^{-1}$) is masked in both fit inputs. Figures for DR5 are very similar, except for including another Gaussian pair for the lower-velocity absorption.

#### 3.3.1. *AOD approach: measuring absorption below continuum using integrated quantities and Apparent Optical Depth method*

Table 2 shows the integrated absorption measurements for both SDSS DR5 and BOSS DR9 epochs. We measured the Balnicity Index (BI; Weymann et al. 1991) modified to the EHVO (BI$_{\rm EHVO}$) by setting the inte-gral limits to account for absorption between 30,000 and 60,000 km s$^{-1}$, but we kept the parameter C to account for absorption larger than 2,000 km s$^{-1}$ (see Paper I for more details). Besides the BI$_{EHVO}$, we also measured the BI using the original definition, setting the integral limits to be 5,000 and 30,000 km s$^{-1}$ (BI$_{orig}$). Table 2 also includes equivalent width values (EW) as the integration of the absorption within the total velocity limits and depth measurements, which were obtained from subtracting 1 minus the local averaged minimum flux value of the trough, as well as upper and lower velocity limits for each EHVO, $v_{max,0.9}$ and $v_{min,0.9}$ respectively, which are determined by the wavelengths where the absorption crosses below the 90% of the normalized flux level. Values in Table 2 are rounded to reflect significant figures. Errors of BI, $v_{max}$, $v_{min}$, EW, and depth are mostly influenced by the location of our continuum fit since it will shift the location of the normalized flux level; we estimated typical errors in Paper I by raising and lowering the normalized continuum by an amount that would place the new continuum fit within the spectrum error (typically by 5% of the normalized flux) and assigning the recalculated measurements as $3\sigma$. We found these errors to be typically in the hundreds of km s$^{-1}$: the BI and EW $\sigma$ errors are less than ∼10% of their values, and typical $v_{min}$ and $v_{max}$ errors are ∼200 km s$^{-1}$; depth errors are ±0.02. All of our absorption measurements are carried out over the unsmoothed spectra, except for the depth value, as the smoothed spectra, where the noise is reduced, are a better representation of the actual real depth.

We find very large values of BI, both for the EHVO and the absorption at lower speeds. Based on the traditional BI definition, J1646 would be considered a BALQSO in the DR5 epoch but not in the DR9 one when this value is zero.

To obtain physical line measurements using the apparent optical depth (AOD) method, we assumed that the line intensities at each velocity $I_v$ are given by

$$I_v = I_o(1 - C_f) + C_f I_o e^{-\tau_v} \tag{1}$$

where $I_o$ is the intensity of the continuum, $C_f$ is the line-of-sight coverage fraction ($0 \leq C_f \leq 1$; see Hamann & Ferland 1999 for more information), and $\tau_v$ is the line optical depth. We assume $\tau_v$ to show a Gaussian profile dependent on the optical depth at the center of the line ($\tau_0$) and the Doppler parameter ($b$). Derived ionic column densities ($N_{ion}$) were calculated as follows

$$N_{ion} = \frac{m_e c}{\pi e^2 f \lambda_0} \int \tau_v dv \tag{2}$$

(Savage & Sembach 1991) where $m_e$ and $e$ are the mass and charge of the electron, respectively, and $f$ is the oscillator strength of the ionic transition at $\lambda_0$. The red line of the doublet is used to avoid potential saturation. Total column density, $N_H$, values are derived from these $N_{ion}$ using a median value for the C IV ion fraction found in photoionization calculations ($f$(C IV) = -1.25, see Appendix in Hamann et al. 2011). We assume solar metallicities (Asplund et al. 2009).

Given the width of the BAL profile in both epochs, the C IV doublet is blended, and the determination of $C_f$, $\tau_0$, and $b$ is an undetermined problem, even more so due to the added possibility of fitting more than one single pair of Gaussian profiles. We explored several tests, allowing the coverage fraction $C_f$ to vary (constrained to be within physical values) or remain fixed, as well as using one doublet or several, always fixing the velocity separation of the C IV doublet and the ratio of optical depths within the doublet (2:1). In our fits, we also masked the region where the C II 1335 Å emission line seems to be present ($-46,500 < v < -44,000$ km s$^{-1}$ in Figure 8), and in DR5 we do not consider the Si IV+O IV] emission line is absorbed.

Table 3 shows a summary of the apparent optical depth (AOD) tests we used to measure the opacity:

- Method 1: Fixing the $C_f$ to be 1 and letting the central velocity $v_0$, the width of the Gaussian $b$, and the central optical depth $\tau_0$ to vary,

- Method 2: An iterative process where in Step (1) we fix $C_f$ to be 1 and let $b$ to vary as before[4], in Step (2) we then use the output $b$ width as a fixed parameter for a fit where $C_f$ are allowed to vary, and lastly, we repeat the first part of the iteration but using the output $C_f$ from the previous step as the fixed parameter.

- Method 3: A similar iterative process as Method 2, but allowing two pairs of Gaussians instead of one, fixing the $C_f$ to be the same for both pairs, i.e., assuming that both gas clouds cover the same region of the quasar (see Rupke et al. 2005).

Table 4 and Figure 8 show the results of these fitting processes. Each line in Table 4 includes the measurement of one Gaussian pair. Figure 8 shows the fitting results for Methods 2 and 3 (the solid purple one) in DR9, where the doublets are indicated as blue/red lines (1548/1550 Å, respectively) – fittings in DR5 were

---

[4] Both $v_0$ and $\tau_0$ are variables through the whole Method 2, and the output of one step is used as the input of the following step.

**Table 2**: Absorption measurements of in J1646

| Epoch | $BI_{orig}$ (km s$^{-1}$) | $BI_{EHVO}$ (km s$^{-1}$) | EW (km s$^{-1}$) | Depth | $v_{min,0.9}$ (km s$^{-1}$) | $v_{max,0.9}$ (km s$^{-1}$) |
|---|---|---|---|---|---|---|
| DR5 | 0 | 6400 | 6800 | 0.72 | -34100 | -50200 |
|  | 2200 |  | 2400 | 0.34 | -15100 | -28300 |
| DR9 | 0 | 3700 | 3900 | 0.61 | -36000 | -49000 |

Typical BI and EW errors are less than ∼10% of their values (approx. hundreds of km s$^{-1}$). Typical $v_{min,0.9}$ and $v_{max,0.9}$ errors are around ∼200 km s$^{-1}$ and for depth errors are ∼ ±0.02. For DR5, the total quasar BI would be the sum of $BI_{orig}$ and $BI_{EHVO}$.

**Table 3**: Summary of AOD fitting methods used in this paper

| Method | Description |
|---|---|
| 1 | Fixing coverage fraction $C_f$ to 1 |
| 2 | Iterative[a] |
| 3 | Iterative[a,b] with two pairs of Gaussians |

[a]Iterative meaning first fixing $C_f = 1$, then fixing $b$ to the output of the previous iteration (all other parameters free), then fixing $C_f$ to the output of the previous iteration (all other parameters free)
[b]Same $C_f$ for all pairs of Gaussians

similar except for adding another Gaussian pair for the lower-velocity absorption. Method 2 resulted in better constrained parameters, meaning smaller error values, for both DR5 and DR9, but the goodness of the fit was, as expected, better with method 3; the standard deviation errors are derived from the covariance matrix of the fit. Both methods are limited by the fact that a single $C_f$ was used for all velocities of the broad profile, and Method 3 has the additional limitation that the same $C_f$ is assumed for both Gaussian pairs used in the C IV EHVO fitting. Given that our goal is to provide a good lower limit for the measurement and a smaller $C_f$ results in a larger $\tau_0$ and therefore a larger column density, we will use the results of method 3 as they provide slightly lower values of column densities.

We find the results are consistent with full or close to full coverage ($C_f \simeq 0.94 - 1.00$), wide Gaussians ($b \simeq 1500 - 7100$ km s$^{-1}$), and optical depths below 1 $\tau \simeq 0.258 - 0.767$). We find lower limits for the column densities log $N_H$ of 21.24 [cm$^{-2}$] and 20.91 [cm$^{-2}$] for DR5 and DR9, respectively. All results included in both tables are lower limits due to several reasons, including: (1) we assumed that the O I and Si IV+ O IV] emission lines are unabsorbed, and thus absorption is restricted to areas below the normalized continuum away from emission, and (2) the measurements are carried out in the spectra normalized by a conservative method (see § 3.1.1).

3.3.2. *Best estimate: absorption properties from SimBAL models*

We extracted the gas physical parameters log $U$, log $N_H$ − log $U$, and the covering fraction parameter log $a$ as well as the gas kinematics (outflow velocity and width) of the BAL outflows from the best-fitting *SimBAL* models. We report the results extracted from the models in which the two spectra were fit simultaneously with gas density fixed at log $n$ = 6 [cm$^{-3}$]. The outflow properties constrained from the *SimBAL* models are tabulated in Table 5. Using the MCMC chains produced by *SimBAL*, we calculated the median values and $2\sigma$ uncertainties from the posterior distributions of each parameter.

The EHVO trough for BOSS DR9 extends from $-34,200$ km s$^{-1}$ to $-48,500$ km s$^{-1}$ with a remarkable velocity width of $v_{width} \sim 14,300$ km s$^{-1}$. The SDSS DR5 revealed even more dramatic BAL that extends to a much lower velocity of $-14,300$ km s$^{-1}$ with $v_{width} \sim 34,200$ km s$^{-1}$. Similar to what was found for DR5 when using the scaled DR9 template (§ 3.1.1; see Figure 2), *SimBAL* also modeled the absorption in DR5 as a single continuous trough; however, the value of the highest velocity end of the trough identified by *SimBAL* is more comparable with the results from the AOD fitting for DR9, but slightly lower for DR5. From the best-fitting models, we obtain the ionization parameters of log $U = -0.7 \pm 0.04$ for the high-velocity portion of the trough that is observed both in DR5 and DR9, and log $U = -1.02^{+0.15}_{-0.13}$ for the lower-velocity part of the BAL, only observed in the DR5 spectrum. We note that all absorption parameters, with the exception of the covering fraction parameter (log $a$), for the high-velocity portion of the trough have been tied to be identical between epochs (§ 3.2.2). Additional *SimBAL* simulations were performed to assess the robustness of the ionization parameter constraints (Appendix B).

We calculated the total outflow column densities log $N_H = 21.79 \pm 0.06$ [cm$^{-2}$] and log $N_H = 21.63^{+0.05}_{-0.06}$

Table 4: C IV absorption AOD fitting results

| Epoch & Method | $v_{max,0.9}$ (km s$^{-1}$) | $v_{centroid}$ (km s$^{-1}$) | $C_f$ | $b$ (km s$^{-1}$) | $\tau_0$ | log N$_{ion}$ [cm$^{-2}$] | log N$_H$ [cm$^{-2}$] | log N$_H$(combined) [cm$^{-2}$] |
|---|---|---|---|---|---|---|---|---|
| DR5 1 | -50200 | -42100 | 1.0 | 5780±150 | 0.79±0.03 | >16.30 | >21.12 | >21.25 |
|  | -28300 | -22800 | 1.0 | 7100±500 | 0.258±0.011 | >15.85 | >20.67 |  |
| DR5 2 | -50200 | -42100 | 0.87±0.06 | 5680±150 | 1.00±0.04 | >16.40 | >21.29 | >21.38 |
|  | -28300 | -22800 | 1.0$^a$ | 7100±500 | 0.258±0.011 | >15.85 | >20.67 |  |
| DR5 3 | -50200 | -47400 | 1.00$^b_{-0.14}$ | 1700±300 | 0.40±0.04 | >16.28 | >21.10 | >21.24 |
|  |  | -40900 | 1.00$^b_{-0.14}$ | 4700±200 | 0.767±0.025 |  |  |  |
|  | -28300 | -22800 | 1.0$^a$ | 7100±500$^b$ | 0.258±0.011 | >15.85 | >20.67 |  |
| DR9 1 | -49000 | -42700 | 1.0 | 4600±100 | 0.60±0.02 | >16.08 | >20.90 |  |
| DR9 2 | -49000 | -42700 | 0.85±0.07 | 4500±100 | 0.79±0.03 | >16.20 | >21.02 |  |
| DR9 3 | -49000 | -47400 | $0.94^{+0.06,b}_{-0.13}$ | 1500±200 | 0.26±0.02 | >16.09 | >20.91 |  |
|  |  | -41600 | $0.94^{+0.06,b}_{-0.13}$ | 3500±100 | 0.69±0.02 |  |  |  |

AOD results of the C IV absorption measurements on the normalized spectra. The methods used in both spectra are described in Table 3. For each method, the results are included from largest to shortest absolute speed (in other words, the EHVO appears first). The best fits were found using method 3 in both epochs.
$^a$ The shape of the absorption at a lower velocity ($v_{max}$ − 28,300 km s$^{-1}$) does not suggest the use of two pairs of Gaussians, and only Method 1 obtains good results. We include the measurement from Method 1 with a single pair.
$^b$ Both $C_f$ are allowed to vary, but they are constrained to have the same value.

Table 5: Absorption measurements extracted from *SimBAL* Models

| Epoch | $v_{max,0.9}$ (km s$^{-1}$) | $v_{min,0.9}$ (km s$^{-1}$) | log $U$ | log $N_H$ − log $U^a$ [cm$^{-2}$] | log $a^{bc}$ | log $N_H{}^c$ [cm$^{-2}$] | log $N_H{}^c$ (combined) [cm$^{-2}$] |
|---|---|---|---|---|---|---|---|
| DR5 | -48500 | -34200 | −0.7 ± 0.04 | 21.36 − 22.95 | 0.71 − 1.2 | $21.71^{+0.05}_{-0.06}$ | 21.79 ± 0.06 |
|  | -34200 | -14300 | $-1.02^{+0.15}_{-0.13}$ | 21.59 − 23.04 | 1.19 − 1.78 | $21.0^{+0.2}_{-0.17}$ |  |
| DR9 | -48500 | -34200 | −0.7 ± 0.04 | 21.36 − 22.95 | 0.9 − 1.3 | $21.63^{+0.05}_{-0.06}$ |  |

$^a$ The range of values estimated from the multiple bins is reported.
$^b$ A large value of log $a$ corresponds to small covering fraction.
$^c$ The total outflow column density is the sum of the covering fraction-weighted column densities calculated for each bin. The parameters as a function of velocity are plotted in Figure 9.

[cm$^{-2}$] for DR5 and DR9, respectively. The total outflow column density is calculated by summing the covering fraction weighted column densities calculated from each tophat bin (Figure 9; log $N_{H,\ corrected}$ = (log $N_H$ − log $U$) + log $U$ − log(1 + 10$^{\log a}$) Arav et al. 2005; Leighly et al. 2018, 2019b; Choi et al. 2022a). We excluded bins with negligible values (log $N_H \lesssim 19.3$ [cm$^{-2}$]) from the total column density calculation, as their uncertainty estimates were bound by the lowest values in the grid.

Figure 9 shows the physical properties of the outflow gas observed in DR5 and DR9 as a function of velocity determined from the best-fitting *SimBAL* models. We found consistently larger values of log $a$ and smaller values of log $N_H$ for all bins in the model for the DR9 spectrum. The figure reports two ionization parameters constrained in the fitting, one for the high-velocity region in both DR5 and DR9 and the other for the lower-velocity region in DR5. We experimented with a set of models where we allowed log $U$ to vary across the velocities; however, they did not produce sufficiently statistically better model fits to justify increasing the number of degrees of freedom and fit parameters of the models. As mentioned in § 3.2.2, gas density was not constrained from the *SimBAL* fitting since there are no absorption lines in the bandpass that are sensitive to the change in density. Although there is neither information in the spectra with which we can estimate the density of the outflow gas nor other EHVO outflows with constrained gas densities, log $n$ = 6 [$cm^{-3}$] is a reasonable assumption for BAL outflow gas, and previous *SimBAL* analysis of a BALQSO SDSS J0850+4451 (Leighly et al. 2018) found similar density constraints. We further discuss the possible range of densities of the outflowing gas in J1646 and resulting wind properties in § 4.2.1.

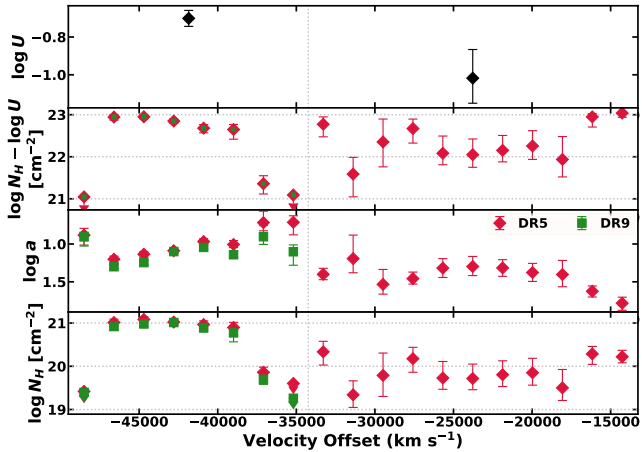

**Figure 9**: The outflow physical properties as a function of velocity from the tophat models for SDSS DR5 and BOSS DR9 spectra. The outflow found in BOSS lacks the lower-velocity part of the outflow seen in the SDSS spectrum. The best-fitting models plotted here used a fixed value of $\log n = 6$ [cm$^{-3}$]. The downward arrows represent the upper-limit estimates because of the finite sizes of the column density grids currently available in *SimBAL*.

## 4. SUMMARY & DISCUSSION

J1646 shows very strong EHVO absorption profiles, resembling some of the strong BALs found at lower speeds. In the next section, we summarize the results found for its extreme outflow, and we discuss the implications of such an energetic outflow in the following ones.

### 4.1. *Summary of Results Obtained with Both Methods*

Using a combination of commonly used conservative methods and state-of-the-art novel spectral synthesis code, we obtain measurements and very-well-constraint lower and upper limits for the absorption parameters.

We found absorption primarily and most strongly in C IV, accompanied by Ly$\alpha$, NV, and Si IV absorption at the same speeds (see Figure 6).

Table 2 includes all speed measurements; typical errors are $\sim$200 km s$^{-1}$. In the first epoch (DR5), the maximum outflow speed reaches $v_{max,0.9} = -50,200$ km s$^{-1}$, slightly decreasing in the second epoch (DR9) to $v_{max,0.9} = -49,000$ km s$^{-1}$. Absorption at lower velocity is only present in the first epoch so the minimum velocity of this outflow dramatically changes from $v_{min,0.9} = -15,100$ km s$^{-1}$ to $v_{min,0.9} = -36,000$ km s$^{-1}$. The location of the Si IV+O IV] emission line in the middle of the absorption profile in DR5 makes it more difficult to attribute the changes in that wavelength region to either the new presence of absorption or changes in the emission.

In this paper, we have presented the results using two different methods. The first method (AOD; § 3.1.1, § 3.2.1, and § 3.3.1) results in a very conservative lower limit of the absorption estimates: it allows us to estimate lower limits of column densities, as smaller coverage fraction $C_f$ values will result in larger optical depths, and hence larger column densities. The second method (*SimBAL*; § 3.1.2, § 3.3.2, and § 3.3.2) provides the best estimate. It allows obtaining very good constraints of ionization $\log U$, the covering fraction parameter $\log a$ that models the inhomogeneous partial covering by using a power-law opacity distribution, as well as a good estimate of the column density.[5]

The absorption in the SDSS DR5 epoch appears overall to be stronger and wider than in the BOSS DR9 one. Using AOD (§ 3.3.1), we measure that $C_f$ might be consistent with full coverage in the high-velocity absorber $1.00_{-0.14}$; we fix it to that value in the measurements of the low-velocity absorber since it cannot be well constrained. Measurements of widths, optical depths, and ionic column densities using this method are included in Table 4 (Method 3). For the EHVO, we measure a column density of $\log N_H > 21.10$ [cm$^{-2}$]. For the lower-velocity absorption, we find a shallower profile with smaller optical depths and a column density of $\log N_H > 20.67$ [cm$^{-2}$]. Overall, we find a lower limit for the column density of the absorption in DR5 of $\log N_H > 21.24$ [cm$^{-2}$]. Using *SimBAL* (§3.3.2), we are able to measure the amount of N V, Ly$\alpha$, and Si IV absorption in the EHVO, which allows us to determine ionization values[6] and the covering fraction parameter (see Table 5). Unlike other EHVO quasars in Paper I, J1646 shows stronger Ly$\alpha$ than N V. Using these measurements, we are able to constrain better the total $N_H$, resulting in a column density of $\log N_H = 21.79 \pm 0.06$ [cm$^{-2}$].

In the BOSS DR9 epoch, the absorption is only present at high speeds and appears slightly weaker over-all. Using AOD (§ 3.3.1), we obtain a $C_f$ that remains consistent with full coverage $0.94^{+0.06}_{-0.13}$ and a total column density of $\log N_H > 20.91$ [cm$^{-2}$] in our best-fitting method (Method 3 in Table 4). When using $SimBAL$ (§ 3.3.2), we tied all absorption parameters between epochs except $\log a$ since it provided the best fitting (see § 3.2.2). Therefore, we find the same ionization for DR9 as in DR5 but different covering fraction parameters over the absorption profile (see Figure 9 for values at each velocity bin). Other ions (N V, Ly$\alpha$, and Si IV) are still present in the same outflow, better constraining the total $N_H$, which results in a column density of $\log N_H = 21.63^{+0.05}_{-0.06}$ [cm$^{-2}$], just slightly lower than in the DR5 epoch.

---

[5] Using the AOD method, we presented a scenario of very broad absorption that would be an upper limit of our estimates: the C IV absorption could extend from $v_{min} = -13,900$ km s$^{-1}$ to $v_{max} = -65,800$ km s$^{-1}$, but it is very uncertain where the potential Si IV or even N V absorption could be confused with C IV absorption. Therefore, we did not include the second scenario in our measurements as it is too unlikely and discouraged by the *SimBAL* method.

[6] In the AOD method, we used a value of $\log U \sim$-0.6, which resulted comparable to the value obtained with *SimBAL* for the EHVO: $\log U \sim$-0.7.

### 4.2. Physical Properties of J1646's Outflow
#### 4.2.1. Location of the Outflow

To estimate the mass outflow rate and the energy the outflow is carrying, we need to determine the location (radius) of the J1646 outflow. This radius ($R$) can be calculated using the definition of the ionization parameter, $U = Q/4\pi R^2 nc$, where the number of photoionizing photons per second emitted from the central engine ($Q$) can be estimated from the analysis of the broadband SED, while the density of the gas ($n$) and the ionization parameter ($U$) are constrained from detailed spectral analysis. We estimated $\log Q = 57.2$ [photons s$^{-1}$] by scaling a standard quasar SED (SED used to create the $SimBAL$ grids; Leighly et al. 2018) to match the J1646 photometry and then integrating the photon flux for energies larger than 13.6 eV. The $SimBAL$ analysis provided us with excellent constraints on the ionization parameters and column densities of the outflowing gas observed in the spectra of J1646. However, due to the lack of density-sensitive diagnostic absorption lines in the spectra, it is not possible to directly constrain the density of the gas. This means we need to infer the location of the outflowing gas using our prior knowledge about BAL cloud geometry and the properties of spectral variability seen in this object.

Many previous studies (e.g., Hamann et al. 1997b; Narayanan et al. 2004; Rodríguez Hidalgo et al. 2011) derive location estimates from the absorption profile variability. Unfortunately, the time interval between the two observations is quite large, which results in a very high upper limit for the distance. If, as those studies do, we assume that the BAL variability is caused by a change in the ionization state of the outflowing gas, we can estimate the density of the gas using the equation, $\Delta t > 1/n_e \alpha_r$, where $\Delta t$ is the BAL variability timescale, $n_e$ is the electron density of the gas ($n$ in $SimBAL$), and $\alpha_r$ is the recombination-rate coefficient (e.g., Hamann et al. 1995; Capellupo et al. 2013). We obtain $n_e > 10^3$ [cm$^{-3}$] with $\alpha_r = 1.5 \times 10^{-11}$ (T= 20000K, McGraw et al. 2017) and $\Delta t \sim 1.7$ years (§ 2). This would put the outflowing gas in this object at $r \lesssim 1500$ pc from the central SMBH. On the other hand, we can obtain a lower limit for the outflow location by considering the gas to be moving transversely; then, we can estimate the distance of the outflows with the equation, $R = \frac{GM_{BH}}{v_{trans}^2}$ where $R$ is the distance of the outflowing gas from the central SMBH, $M_{BH}$ is the mass of the SMBH, $G$ is the gravitational constant, and $v_{trans}$ is the transverse velocity of the clouds that can be approximated as Keplerian speed (Capellupo et al. 2011; Moravec et al. 2017). Although it is possible to derive an observation-based transverse velocity based on variability timescales, this method requires a higher cadence spectroscopic campaign to estimate a reliable transverse velocity (ideally more than two epochs that capture both the appearance and disappearance of the trough), which does not exist for J1646, as we mentioned above. We take the mass of the central SMBH in J1646 $\log M_{BH} \sim 9.5$ M$_\odot$ (Rankine et al. 2020) and assume that the LOS outflow velocity we measure from the shift of the BALs as the plausible upper limit of the transverse velocity $v_{trans} \lesssim 50,000$ km s$^{-1}$ (e.g., Hall et al. 2011). These values return $R \sim 0.006$ pc, which is about 18 times the Schwarzschild radius ($R_s \sim 0.0003$ pc).

To derive better upper and lower limits on the radius of the wind in J1646, we made assumptions about the energetics and physical properties of the EHVO winds based on our prior knowledge about BAL winds following the approach introduced in Bischetti et al. (2024). We estimated the lower limits of the BAL radius to be $R \gtrsim 5 - 7$ pc conservatively assuming that a typical BAL outflowing gas has $\log n \lesssim 8$ [cm$^{-3}$], consistent with the range of gas densities found in previous work with $SimBAL$ (Choi et al. 2022b; Green et al. 2023; Leighly et al. 2018) and using the equation for the ionization parameter above. However, we note that higher gas densities ($\log n \gtrsim 9$ [cm$^{-3}$]) have been reported in a few extreme BAL systems with Balmer line absorption (e.g., Hall et al. 2007; Schulze et al. 2018).[7] The lower range value was calculated for the EHVO trough that was constrained to have a higher-ionization parameter ($\Delta \log U \sim 0.3$) than the disappearing low-velocity trough.

---

[7] There are several scenarios where the distance-density gas relationship would be more complex; for example, self-shielded gas where different strata of the gas receive a different influx of central radiation or patchy/compacted clouds of gas.

The upper limit of the BAL radius was derived assuming that a steady BAL wind, which may be radiatively driven (e.g., King & Pounds 2015; Zubovas & King 2012), cannot transport energy greater than the radiative power of the quasar. The kinetic luminosity or power of the outflow is given by the equation, $L_{KE} = \dot{M}_{out} v_{out}^2/2$. The mass outflow rate ($\dot{M}_{out}$) can be calculated using the equation, $\dot{M}_{out} = 8\pi \mu m_p \Omega R N_H v_{outflow}$ (Dunn et al. 2010) where the mean molecular weight is assumed to be $\mu = 1.4$. The global covering fraction ($\Omega$) is estimated from the population statistics of typical (non-EHVO) BAL quasars. Assuming an expanding shell geometry of BAL winds, the fraction of outflowing gas covering the entire solid angle is expected to correspond to the fraction of the quasar population that shows BAL features. Given the lack of information about the EHVO-quasars fraction among the quasar population, an accurate estimate of the EHVO global covering fraction is not available. Therefore, we assumed that the EHVOs share a similar outflow geometry as non-EHVOs and used the $\Omega = 0.2$, typically adopted for BAL winds (e.g., Hewett & Foltz 2003). A smaller covering fraction would result in a lower value for the mass outflow rate proportionally. Notice that while the kinetic luminosity is proportional to the radius of the outflow, it is more strongly dependent on the velocity of the outflow, as it is proportional to $v^3$.

We estimated the upper limits of the outflow radius for the EHVO at $R \lesssim 30$ pc, the values at which the kinetic luminosity of the outflow approaches the quasar luminosity $L_{KE} \sim L_{Bol}$ for DR5 and DR9, respectively. The small difference in the plausible upper limits is due to the lower outflow column density in DR9 BAL. For the lower-velocity BAL, a significantly larger upper limit of $R \lesssim 540$ pc can be assumed, as the lower outflow velocity permits $R$ to reach higher values before the kinetic luminosity matches the quasar luminosity. These upper limits correspond to the gas density of $\log n \gtrsim 4.2 - 6.5$ [cm$^{-3}$]. Figure 10 summarizes the process by which we obtained the plausible range of radii of the outflows seen in J1646.

We conservatively estimate that the outflowing gas for the EHVO, which we observe in both epochs, is likely located at $5 \lesssim R \lesssim 30$ pc, and the low-velocity BAL, only present in DR5, is likely at $7 \lesssim R \lesssim 540$ pc away from the central engine. It can be seen that a much smaller range of $R$ is obtained for the EHVO due to its high outflow velocity, restricting the plausible location of the outflow gas much closer to the central SMBH. We stress that our results do not explicitly preclude the possibility that the gas clouds creating the high-velocity and low-velocity troughs could be at the same location in the quasar. The difference in constrained $\log U$ values between the high- and low-velocity troughs is marginal, given their uncertainties. Furthermore, we conducted an additional suite of $SimBAL$ simulations using a single $\log U$ parameter to fit the entire trough observed in DR5. The results indicated that a $\log U \sim -0.7$ provided a satisfactory fit across both the EHVO and low-velocity BAL regions, consistent with the constraints extracted from the best-fitting model for the EHVO trough. In other words, it is possible that a single gas cloud (or a group of cloudlets) produced the wide BAL features observed in DR5, and the change in geometry or spatial distribution of the gas relative to the LOS caused the lower part of the BAL trough to be absent in DR9. We discuss in detail the potential scenarios for the variability in § 4.3.3.

Even higher (lower) density will place the gas unrealistically too close (far) from the central SMBH, which has not been observed in BAL quasars (e.g., Leighly et al. 2018; Choi et al. 2022b,a). Robust detections of outflows within the broad line region (BLR) remain elusive. While large radii of BAL winds have been discovered (e.g., $R \sim$ 10s kpc; Byun et al. 2022), the extremely high outflow velocities found in J1646 disfavor such a large-scale expanding shell scenario, as it would result in the unrealistically high kinetic luminosity of the wind and challenge current models of wind acceleration at large radii (e.g., Faucher-Giguère et al. 2012; Faucher-Giguère & Quataert 2012).

Although it is difficult to further narrow the range of expected locations of the outflowing gas, we can assume that the actual location of the outflowing gas may be closer to the lower bound of the range. This is because we observe unabsorbed emission lines present within the absorption line profile: Si IV+O IV] in the DR5 epoch, resembling the shape of the emission lines in the later epoch, and C II in both epochs. We showed that the emission lines, in general, are very similar in both epochs (see § 3.1.1), and our best-fitting $SimBAL$ model included them also unabsorbed (§ 3.1.2). Such differential partial covering between line emission and continuum emission is difficult to produce with gas clouds located at a large distance because the angular sizes of the emission sources become extremely small (e.g., Leighly et al. 2019b; Choi et al. 2022a), even if the gas is porous/patchy and the emission is not in our line of sight. However, this possibility cannot be ruled out, as it is well established that the BLR does not fully cover the central SMBH ($\Omega_{BLR} \lesssim 0.2$; e.g., Leighly 2004); otherwise, all X-ray spectra would be significantly absorbed. BAL outflowing gas also covers a portion of the central engine ($\Omega_{BAL} \sim 0.2$; e.g., Hewett & Foltz 2003), though there is no requirement that $\Omega_{BLR}$ and $\Omega_{BAL}$ cover the same solid angle. Similar differential partial covering, in which line emission appeared unabsorbed by BAL outflows, has been reported in several FeLoBAL quasars (e.g., Shi et al. 2016; Choi et al. 2022b) with similar bolometric luminosities ($\log L_{Bol} \sim 46 - 47$ [erg s$^{-1}$]). They reported that the BAL outflowing gas is located at distances ranging from approximately $\sim$ pc to $\sim 10$ pc from the central SMBH, which are greater than the assumed size scales of the BLR ($R_{BLR} \lesssim$ pc). Such residual emission features at the bottom of BAL troughs are not uncommon (e.g., Wampler et al. 1995) and are likely caused by multiple factors, including partial coverage of the BLR by BAL gas (e.g., Leighly et al. 2019b) and scattering of BLR emission by dust (e.g., Choi et al. 2020; Ogle et al. 1999).

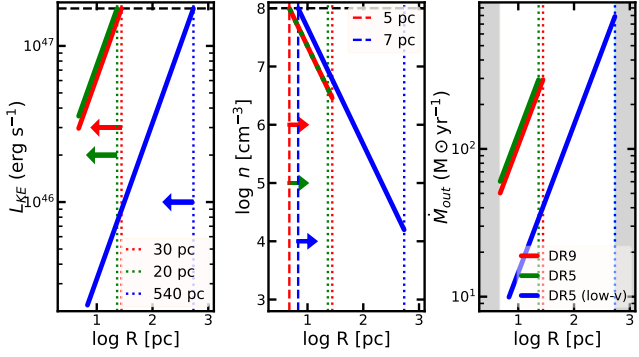

**Figure 10**: *Left panel:* The kinetic luminosity of the outflow as a function of the plausible range of distances of the outflow from the central SMBH. The dashed horizontal black line represents $L_{Bol}$, which we considered as the upper limit for outflow power. The arrows represent the upper limits on $\log R$, which were set by assuming $L_{KE} \lesssim L_{Bol}$ for DR5 (low-v; blue), DR5 (EHVO, green) and DR9 (EHVO, red). *Middle panel:* The range of gas densities corresponds to the range of distances of the outflow from the central SMBH. The dashed horizontal black line represents the upper limit on gas density, $\log n \lesssim 8$ [cm$^{-3}$], that we assumed to estimate the lower limits on the radii of BAL outflows, which are shown as vertical dashed lines and arrows. *Right panel:* The mass outflow rates of the BAL outflows as a function of the plausible radii of BAL outflows.

The radius of the BLR can be estimated from the monochromatic flux at 5100 Å (Bentz et al. 2013) and appears to be around $\log R_{BLR} \sim 0.83$ pc for J1646. While we measure radial speeds, if we assume a correlation between the radial component and the total outflow velocity, it would be reasonable to conclude that EHVOs must be closer to the source than other typical BAL outflows at lower speeds. In fact, the best estimates of the location of Ultra Fast Outflows relative to the central source place them at a distance of $5R_s$. For all these reasons, we expect the BAL outflow in J1646 to be located tens of parsecs from the central SMBH.

If the outflow in J1646 is located close to the central SMBH, Figure 10 shows that the density should be quite high for all absorbers. A higher density leads to a lower volume filling factor, presenting a greater challenge for explaining BAL cloud confinement (e.g., Choi et al. 2022a). For a fixed density, the EHVOs show a slightly smaller distance ($R$) than the lower-velocity absorber, so we might be seeing different parts of the outflow path.

Using the nominal speed of $-50,000$ km s$^{-1}$, the outflow would have traveled just $\sim 0.09$ pc in the quasar rest-frame between the two epochs, so we can assume a similar location in the two epochs, provided the cloud is long lasting.

#### 4.2.2. *Mass, Energy, and Acceleration Mechanism of the Outflow*

In § 4.2.1, we calculated the mass outflow rates and the kinetic luminosities of the outflows using the properties constrained by the *SimBAL* analysis to derive a plausible range of location of the outflows (Figure 10). For the EHVO troughs that are present in both epochs, we estimated the mass outflow rate $\dot{M}_{out} \sim 60 - 290$ M$_\odot$ yr$^{-1}$ and $\dot{M}_{out} \sim 50 - 290$ M$_\odot$ yr$^{-1}$ for DR5 and DR9, respectively, corresponding to the range of $5 \lesssim R \lesssim 30$ pc. The range of mass outflow rates for the BAL in DR9 extends to slightly lower values due to the lower column density (Table 5). We estimated the kinetic luminosity of $\log L_{KE} \sim 46.5 - 47.2$ [erg s$^{-1}$].

The low-velocity component is estimated to have a mass outflow rate of $\dot{M}_{out} \sim 10 - 790$ M$_\odot$ yr$^{-1}$ and a kinetic luminosity of $\log L_{KE} \sim 45.3 - 47.2$ [erg s$^{-1}$]. As mentioned above, we assumed the upper limit for the kinetic luminosity as the bolometric luminosity of the quasar, $\log L_{Bol} = 47.2$ [erg s$^{-1}$]. If the outflowing gas clouds for the two velocity components are located co-spatially, the bulk of the outflow energy and mass would be mainly transported by the high-velocity component, given that the outflow strengths strongly depend on the outflow velocity, $L_{KE} \propto v_{outflow}^3$. We reiterate that the kinetic luminosity and the mass outflow rate of the outflow are both proportional to the assumed location of the outflow because we assume an expanding shell geometry. Combining both the low-velocity BAL and the EHVO components in DR5, assuming that they are physically independent of each other, we estimate a total mass outflow rate of $\dot{M}_{out} \sim 70 - 1080$ M$_\odot$ yr$^{-1}$ and $\log L_{KE} \sim 46.6 - 47.5$ [erg s$^{-1}$] for the total outflow power.

Table 6: Outflow properties calculated from *SimBAL* Results

| Epoch & BAL | $R$ (pc) | $\dot{M}_{out}$ ($M_\odot$ yr$^{-1}$) | $\dot{M}_{out,\ total}$ ($M_\odot$ yr$^{-1}$) | log $L_{KE}$[a] [erg s$^{-1}$] | log $L_{KE,\ total}$ [erg s$^{-1}$] |
|---|---|---|---|---|---|
| DR5, EHVO | $5-20$ | $60-290$ | $70-1080$ | $46.6-47.2$ | $46.6-47.5$ |
| DR5, low-$v$ BAL | $7-540$ | $10-790$ | | $45.3-47.2$ | |
| DR9, EHVO | $5-30$ | $50-290$ | | $46.5-47.2$ | |

[a] The upper limits for the outflow kinetic luminosities were estimated to not exceed the quasar luminosity ($L_{KE} \lesssim L_{Bol}$; see § 4.2.1 and Figure 10).

The momentum flux ratio ($\dot{P}_{outflow}/\dot{P}_{AGN}$; $\dot{P}_{outflow} = \dot{M}_{out}v_{outflow}$, $\dot{P}_{AGN} = L_{Bol}/c$) can be used to infer potential outflow acceleration mechanisms of quasar outflows. For instance, $\dot{P}_{outflow}/\dot{P}_{AGN} \sim 1$ is expected for radiatively driven winds in which the outflowing gas is accelerated by the scattering of photons (momentum conserving; e.g., King 2003; King & Pounds 2003). The ratio can exceed unity when dust is present in the gas cloud (Thompson et al. 2015). Quasar outflows with $\dot{P}_{outflow}/\dot{P}_{AGN} \gg 1$ are also observed in ionized winds and molecular outflows (e.g., Fiore et al. 2017). These outflows are believed to be powered by an energy-conserving acceleration mechanism and are found mainly at large distances (log $R > 2$ [pc]) from the central SMBH (e.g., King & Pounds 2015). For J1646, we estimate $3 \lesssim \dot{P}_{outflow}/\dot{P}_{AGN} \lesssim 15$ and $0.3 \lesssim \dot{P}_{outflow}/\dot{P}_{AGN} \lesssim 24$ for the high- and low-velocity troughs, respectively. Although these values are greater than unity, they are consistent with what has been observed in (FeLo)BAL outflows (e.g., Fiore et al. 2017; Choi et al. 2022b). While these simple calculations revealed no compelling observational evidence that special acceleration mechanisms are required to explain the EHVO's extremely high velocity, further theoretical investigation is needed to better understand their acceleration.

### 4.3. Setting J1646 and its Outflow in the Context of Other Quasars

#### 4.3.1. Comparison to Other EHVOs and EHVO quasars and Other BALs and BALQSOs

The EHVO absorption feature in J1646 is rare in the sense that absorption observed at these speeds in UV/optical quasar spectra is not typically so wide and strong, showing such large equivalent widths and depths. In Paper I, we found that, among the 6760 quasars sampled, none of the other 47 EHVO profiles discovered had equivalent widths or depths larger than the absorption feature found in the J1646 spectra. The second strongest profiles showed only 37% (EW = 2500 km s$^{-1}$) or 80% (depth = 0.57) of the properties found in the BOSS DR5 spectrum of J1646, with most values much lower than 50%. J1646 also shows the largest width ($\Delta v \sim$16,100 km s$^{-1}$). Similarly, other EHVO quasars in the literature (e.g., Jannuzi et al. 1996; Rodríguez Hidalgo et al. 2011) were observed to show weaker and narrower absorption. As a consequence, J1646 shows a larger $N_{ion}$ by more than one order of magnitude than the value found by Rodríguez Hidalgo et al. (2011) of $N_{ion} \sim$0.8 $\times 10^{15}$ cm$^{-2}$ for the C IV EHVO in PG0935+417.

Compared to measurements of other BALs (i.e., Filiz Ak et al. 2013), EHVOs show weaker troughs. Thus, it is natural that they tend to show relatively lower column density.[8] Typical values of column densities found in BALQSOs are in the range $19 \lesssim \log N_H \lesssim 23$ [cm$^{-2}$] (Leighly et al. 2018, Choi et al. 2022b, Green et al. 2023 Bischetti et al. 2024, Byun et al. 2024), with FeLoBALs and LoBALs, showing larger values in general (Hamann et al. 2019). Compared to these studies, the column density of the wind in J1646, log $N_H = 21.6-21.8$ [cm$^{-2}$], is relatively low, largely consistent with its classification as a HiBAL, but we advise caution when comparing different classes, as the general properties of EHVO outflowing gas and HiBAL outflows remain poorly understood.

While the column densities for EHVOs are lower than those found in other BAL classes, their extreme speeds allow them to drive high-mass outflow rates and large kinetic luminosities. Indeed, the EHVO in J1646 represents one of the most powerful quasar outflows observed to date with log $L_{KE} \sim 46.5 - 47.2$ [erg s$^{-1}$] ($L_{KE}/L_{Bol} \gtrsim 0.18$). For comparison, Choi et al. (2020) and Choi et al. (2022b) reported log $L_{KE} \sim$ 48.1 [erg s$^{-1}$] and a highest value of log $L_{KE} \sim$ 46.5 [erg s$^{-1}$] (which corresponded to $L_{KE}/L_{Bol} \sim$ 0.18) for a $v_{BAL} \sim -38,000$ km s$^{-1}$ wind found in a luminous FeLoBAL quasar at $z \sim 2.26$ and a sample of low-redshift ($z \sim 1$) FeLoBAL quasars analyzed using

---

[8] Note that J1646 was classified as a BALQSO based on the C IV absorption at lower speeds in DR5 (Gibson et al. 2009a). This absorption disappears in the DR9 observation, while the EHVO remains. As we have cautioned before, this classification based on one-time observations might not be optimal, as both EHVOs and BALs are known to be highly variable.

*SimBAL*, respectively. Bischetti et al. (2024) analyzed a high-redshift LoBAL quasar at $z \sim 6.6$ using *SimBAL* to constrain an energetic wind with $\log L_{KE} \sim 45.7 - 47.5$ [erg s$^{-1}$]. This underscores the potential role of EHVOs in powering quasar feedback.

Even more, the measurements in J1646 might be underestimated. Several possibilities might be causing EHVOs to appear as weaker and shallower absorption profiles: (1) a small covering fraction, which is justified by (a) a Keplerian model where the fastest velocities would be observed closer to the SMBH, (b) our location estimates (see § 4.2.1 and Figure 10), and (c) the large values of $\log a$ obtained through the absorption profile with *SimBAL*[9] fitting (see Figure 9); (2) incident spectral energy reduced by relativistic effects, which results in shallower profiles as the outflowing speeds increase (see Luminari et al. 2020). We describe these more in depth in our survey of EHVO quasars in DR16 (Rodríguez Hidalgo et al. in prep).

In Paper I, we found that our sample of EHVO quasars in DR9 showed larger bolometric luminosities overall than the BALQSOs and non-BALQSOs in our parent sample. We have confirmed this result for the EHVO quasars in DR16 (Rodríguez Hidalgo et al in prep). However, among EHVO quasars alone, the properties of J1646 do not seem to make it an outlier, showing both a large bolometric luminosity and a large black hole mass, but a median value of the Eddington ratio. J1646 is a luminous quasar ($g$ = 18.80 at MJD = 52760). EHVO quasars appear to show a distribution of larger values of bolometric luminosities compared to the parent sample, and J1646 lies within the top third or second quartile of other EHVO quasars in the Paper I sample, depending on the measurement.[10] Rankine et al. (2020) measured a black hole mass of $\log(M_{BH}/M_\odot =)$ 9.5,[11] and an Eddington ratio of $\log(L_{bol}/L_{Edd})$ = -0.4. The value of black hole mass situates it in the top quartile (median = 9.39, range = 8.90 – 9.58), but the Eddington ratio is very close to the median (median = -0.38, range = -0.72 – 0.09).

J1646 is also not an outlier relative to other quasars with EHVOs based on the properties of its C IV emission line. Rodríguez Hidalgo & Rankine (2022) carried out a study of the C IV emission-line parameter space for EHVOs in relation to BALQSOs and non-BALQSOs, finding that EHVOs show overall much larger values of C IV blueshift than the other two samples; this result has been confirmed when investigating a larger sample of EHVOs (Flores et al. in prep). Among the EHVOs in Rodríguez Hidalgo & Rankine (2022), J1646 shows values close to median values with a C IV emission blueshift of $\sim$2470 km s$^{-1}$ and a log(EW) of the C IV emission line of 1.334 Å. An additional way of representing this uses the value of the C IV distance (Rivera et al. 2020); for J1646 this value is 0.85, which also lies close to the median value of the EHVO distribution.

The strength of the He II $\lambda$1640.42 emission line can be used as a proxy for the presence of soft X-ray continuum emission; Casebeer et al. (2006) showed that the He II emission line is theoretically stronger for harder SEDs. Harder SEDs may overionize the gas, making it less likely to observe strong winds. Composites of BAL-type quasars have typically weaker He II emission (Richards et al. 2011). In Paper I, the EW of the He II emission line was measured in all of the EHVO quasars, following a similar procedure to Baskin et al. (2013) and Baskin et al. (2015). In J1646, He II is embedded within the plateau of Fe II emission lines redward of the C IV emission line, and is only clearly evident in the DR9 spectrum. Between the two epochs in J1646, we find that the absorption is weaker when the He II emission appears stronger, as expected.

Finally, another way in which J1646 seems to differ from other EHVOs in our preliminary work (Paper I) is in the presence of Ly$\alpha$ as part of the outflow. J1646 is the only case in which Ly$\alpha$ seems similar or even stronger than N V. This result seems to be confirmed by the *SimBAL* fitting (see Fig.6). In Paper I, only five cases showed absorption at the wavelengths that would indicate Ly$\alpha$, while 26 cases showed potential N V. Together with the presence of Si IV, it indicates this case is more optically thick than other EHVOs in our sample.

---

[9] Increasing $\log a$ corresponds to decreasing $C_f$ (see, e.g., Leighly et al. 2019b). However, we note that inhomogeneous partial covering by $\log a$ and the homogenous (step-function) partial coverage using $C_f$ assume different BAL cloud geometry, and therefore, a direct comparison between the two is not straightforward, as they represent distinct assumptions about the distribution and structure of absorbing material along the line of sight (e.g., Leighly et al. 2019b). Moreover, the values of $\log a$ were constrained by *SimBAL* using multiple BAL transitions, while $C_f$ was derived by AOD solely from the C IV BAL trough. The two methods coincide on the direction of the change, as both of them show a reduction in partial coverage from measurements on DR5 to DR9.

[10] The value calculated by Shen et al. (2011) for J1646 is $M_i[z=2]$ is -28.9 (MJD = 53167), which lies in the top third of values of the 21 quasars with EHVO that we studied in Paper I The bolometric luminosity measured by Rankine et al. (2020) of $log(L_{bol}/[\text{erg s}^{-1}])$ = 47.25 lies in the second quartile of values (median = 47.07, min = 46.58 and max = 47.55).

[11] Values of $M_{BH}$ included in Paper I from Shen et al. (2011) are overestimated, which also affects the Eddington ratios.

### 4.3.2. *Comparison to UFOs and X-Ray Observations*

EHVOs are not the only outflows detected with such extreme speeds. Ultra-fast outflows (UFOs) have been observed as Fe K-shell absorption in the X-ray spectra of nearby AGN (predominantly Seyferts) at similar and even higher speeds (0.03c - 0.4c; e.g., Chartas et al. 2002; Tombesi et al. 2010; and references therein). Because UFOs are searched for, mostly, in local AGN ($z_{em} < 0.1$), the central engine shows typically moderate luminosities, $\log L_{\rm Bol} \sim 43$ - 45.5 [erg s$^{-1}$] (this was calculated as $L_{\rm Bol} = k_{Bol} L_{ion}$ assuming a k$_{\rm Bol}$ of 10, see Tombesi et al. 2013). The upper limit of this range is still two orders of magnitude smaller than the value Rankine et al. (2020) found for J1646 ($\log L_{\rm Bol} = 47.25$ [erg s$^{-1}$]). The detection of UFOs in high-z quasars is still rare, but if the trend between velocity and $L_{\rm Bol}$ is extended to the X-ray absorbers, it might imply that UFOs in quasars with larger luminosities would show even larger velocities.

While we present our best value of column density and utilize it for the calculation of mass outflow rate, and thus, outflow power, there are several indications that these results might still be underestimated. First, values of $N_H$ might be slightly underestimated in both methods, as we have not taken into account relativistic effects, as we discuss in section 4.3.1. Second, there is a very limited number of studies of extremely high velocity outflows observed in X-rays due to the redshifts of the quasars in which they are detected. In Sabra et al. (2003), a study of the quasar PG 2302+029, which presented an EHVO outflowing at ∼56,000 km s$^{-1}$, showed that the column densities derived from the UV absorption lines did not surpass our values – their largest $\log N_{ion}$ was 15.7 [cm$^{-2}$] derived from O VI, the $\log N_{ion}$ measured for C IV was 14.9 cm$^{-2}$. However, when observed in X-rays, they measured a column density from a related outflowing absorber that totaled $\log N_H = 22.4$ [cm$^{-2}$]. If we also had an accompanying UFO, a seven order magnitude increase from our values would result in a column density of $\log N_H = 23 - 24$ [cm$^{-2}$].

In summary, assuming similar effects for UV/optical absorption as the effects in X-rays and that EHVOs are accompanied by ultra-fast X-ray absorbers, our results would be largely underestimated.

### 4.3.3. *Comparison to other SimBAL Variability Analyses*

The simultaneous fitting using *SimBAL* of multiple epochs can provide information about the physical origin of the absorption variability. Because *SimBAL* allows the study of physical parameters for BAL outflows in great detail (e.g., physical properties as a function of velocity), we can investigate which physical change in the gas is responsible for the observed BAL variability. This analysis can be performed by investigating which parameter or a set of parameters can best model the spectroscopic time series data.

Green et al. (2023) used *SimBAL* in this manner for WPVS 007. They analyzed four epochs of HST/COS spectra, revealing that changes in the covering fraction of the outflow gas ($\log a$) were the primary driver of variability, rather than changes in ionization parameter or column density of the outflow. The key findings in their work indicate that this variability likely originated from structural changes within the outflowing gas itself rather than a transverse motion of the outflow gas, a common interpretation for BAL variability due to changes in partial coverage. They argued that changes in covering fractions are caused by the formation and dissipation of clumps of material along the line of sight with the use of $\log a$ in *SimBAL*, which models inhomogeneous power-law partial covering (Leighly et al. 2018; de Kool et al. 2002; Sabra & Hamann 2005). This representation of partial coverage can be interpreted as clumpy structures within the BAL outflow cloudlets (Leighly et al. 2019b), so that the changes in power-law partial covering may be interpreted as changes in the diffuseness of these clumps along our line of sight.

The multi-epoch spectra of J1646 were analyzed in a similar way to the approach taken by Green et al. (2023). As discussed in § 3.2.2, we simultaneously fit the two epochs with *SimBAL* and allowed only a single parameter to vary to isolate the primary source of variability. This analysis revealed that the model that allowed covering fraction parameters to vary produced the best fits to the J1646 spectra, which is the identical model used for WPVS 007. Such results suggest that the EHVO variability seen in J1646 is also potentially caused by the changes in the sub-structures of the cloudlets that produce the absorption features (see Figure 11 in Green et al. 2023).

## 5. CONCLUSIONS

We perform the analysis of two-epoch spectra for the quasar J1646 ($z_{em}$ ∼3.04), to characterize the properties of its EHVO, observed in C IV, Si IV, N V and Ly$\alpha$ and remarkably variable between the two epochs separated ∼ 1.7 years (rest-frame), and the disappearing absorption at lower speeds. We approach the analysis in two ways: first, we employ a traditional method using AOD measurements, and second, we use *SimBAL* to perform simultaneous spectral fits. Here, we summarize the key results:

1. J1646 shows the widest and strongest absorption found in an EHVO to date. Its SDSS DR5 spec-

24trum reveals a wide C IV BAL trough spanning from $-50,200$ km s$^{-1}$ to $-15,100$ km s$^{-1}$. The BOSS DR9 spectrum shows a dramatic change in the C IV BAL, where the depth of the EHVO trough became weaker, and the range of velocities changed to $-49,000$ km s$^{-1}$ – $-36,000$ km s$^{-1}$; this dramatic change at the lower end is due to the fact that the lower-velocity BAL disappeared altogether. J1646 also shows the only case in Paper I in which the absorption in Ly$\alpha$ seems similar or even stronger than N V. Together with the presence of Si IV, it indicates this case might be more optically thick, but analysis of the rest of the sample is necessary.

2. Using the conservative AOD approach described in § 3.3.1, we measure optical depths for the EHVO of $\tau_0 = 0.767\pm0.025$ and $\tau_0 = 0.40\pm0.04$, and we find a coverage fraction $C_f$ consistent with full coverage, which combined result on a lower limit for the ionic column density of $\log N_{\rm ion} > 16.28$ [cm$^{-2}$], and a column density of $\log N_{\rm H} > 21.10$ [cm$^{-2}$]. For the low-velocity absorption, we find a shallower profile with smaller optical depths of $\tau_0 \sim 0.258\pm0.011$, an $\log N_{\rm ion} > 15.85$ [cm$^{-2}$], and a column density of $\log N_{\rm H} > 20.67$ [cm$^{-2}$]. Overall, we find a lower limit for the column density of the absorption in DR5 of $\log N_{\rm H} > 21.24$ [cm$^{-2}$]. In the BOSS DR9 epoch, we obtain a $C_f$ that remains consistent with full coverage ($0.94^{+0.06}_{-0.13}$), smaller optical depths of $\tau_0 \sim 0.69\pm0.02$ and $\tau_0 \sim 0.26\pm0.02$, a $\log N_{\rm ion} > 16.09$ [cm$^{-2}$], and a total column density of $\log N_{\rm H} > 20.91$ [cm$^{-2}$]. These results appear to be accurate lower limits relative to the results obtained with *SimBAL* below.

3. We use *SimBAL* to simultaneously fit the multi-epoch spectra to constrain the outflow properties and investigate the physical origin of the variability. We obtain a best estimate for the outflow column density of $\log N_H = 21.79 \pm 0.06$ [cm$^{-2}$] for DR5 and $\log N_H = 21.63^{+0.05}_{-0.06}$ [cm$^{-2}$] for DR9. We find an ionization parameter $\log U = -0.7 \pm 0.04$ for the EHVO in both epochs and $\log U = -1.02^{+0.15}_{-0.13}$ for the lower-velocity BAL that is only present in DR5 (§ 3.3.2).

4. We estimate plausible ranges for the location of the outflow at $5 \lesssim R \lesssim 28$ pc for the EHVO and $7 \lesssim R \lesssim 540$ pc for the lower-velocity BAL absorber (see § 4.2.1). Based on these estimates, we calculate the outflow power as $\log L_{KE} \sim 46.5 - 47.2$ [erg s$^{-1}$] for the EHVO and $\log L_{KE} \sim 45.3 - 47.2$ [erg s$^{-1}$] for the lower-velocity BAL, as well as mass outflow rates of $\dot{M}_{out} \sim 60 - 290$ M$_\odot$ yr$^{-1}$ in the first and $\dot{M}_{out} \sim 50 - 290$ M$_\odot$ yr$^{-1}$ in the second epoch for the EHVO, and $\dot{M}_{out} \sim 10 - 790$ M$_\odot$ yr$^{-1}$ for the lower-velocity BAL. The EHVO may be driving most of the mass outflow rate and kinetic luminosity, especially if the EHVO is cospatial or at larger distances than the lower-velocity outflow. Combined with its extreme speeds and the observation that several of these measurements might be underestimated, the EHVO in J1646 represents one of the most powerful quasar outflows observed to date, even when compared to LoBALs and FeLoBALs.

5. While EHVO quasars as a whole show distinct physical properties relative to BALQSOs and non-BALQSOs, we find that J1646 is not an outlier among other EHVO quasars (see § 4.3.1). Its Eddington ratio ($\log L_{\rm bol}/L_{\rm Edd} = -0.4$) is close to the median value of EHVO quasars, and its bolometric luminosity ($\log L_{\rm bol} = 47.25$ [erg s$^{-1}$]) is in the second quartile. Only its black hole mass ($\log M_{\rm BH} = 9.5$ [$M_\odot$]) is in the top quartile of those studied in Paper I. J1646 is also not an outlier in the C IV parameter space.

6. We investigate the potential origin of the dramatic variability observed between the two epochs in § 4.3.3. The simultaneous multi-epoch fitting with *SimBAL* revealed that the changes in the covering fraction were the primary driver of BAL variability observed in J1646. This result is consistent with what has been found in the *SimBAL* analysis of WPVS 007 reported in Green et al. (2023).

All of these results underscore the crucial role of EHVOs in powering quasar feedback, the need to account for the energy carried by these outflows, and that further study is necessary to understand this type of quasars. Future surveys, such as SDSS-V, will provide key long-baseline spectral time series data that will further help us identify and study new variable EHVO quasars in greater detail.

P.R.H., M.M.C., and L.F. acknowledge support from the National Science Foundation AAG Award AST-2107960, funding provided by UW Bothell, and from the UW Royalty Research Fund A162630, as well as the Mary Gates Endowment through their research scholarships. Their work was partially supported by the Sloan Digital Sky Survey's Faculty And Student Team program, funded by the Alfred P. Sloan Foundation. H.C. and K.K.L. acknowledge support from the National Science Foundation AAG award Nos. AST-1518382, AST-2007023, and AST-2006771 to the University of Oklahoma. Support for *SimBAL* development and analysis is provided by NSF Astronomy and Astrophysics grants No. 1518382, 2006771, and 2007023. P.B.H. acknowledges support from NSERC grant 2023-05068. Funding for SDSS-III has been provided by the Alfred P. Sloan Foundation, the Participating Institutions, the National Science Foundation, and the U.S. Department of Energy Office of Science. The SDSS-III web site is http://www.sdss3.org/. SDSS-III is managed by the Astrophysical Research Consortium for the Participating Institutions of the SDSS-III Collaboration including the University of Arizona, the Brazilian Participation Group, Brookhaven National Laboratory, Carnegie Mellon University, University of Florida, the French Participation Group, the German Participation Group, Harvard University, the Instituto de Astrofisica de Canarias, the Michigan State/Notre Dame/JINA Participation Group, Johns Hopkins University, Lawrence Berkeley National Laboratory, Max Planck Institute for Astrophysics, Max Planck Institute for Extraterrestrial Physics, New Mexico State University, New York University, Ohio State University, Pennsylvania State University, University of Portsmouth, Princeton University, the Spanish Participation Group, University of Tokyo, University of Utah, Vanderbilt University, University of Virginia, University of Washington, and Yale University. Some of the computing for this project was performed at the OU Supercomputing Center for Education & Research (OSCER) at the University of Oklahoma (OU). JHL is funded through the Canada Research Chair program, as well as the Natural Sciences and Engineering Research Council of Canada (NSERC) Discovery and accelerator grant programs. The Université de Montréal recognizes that it is located on unceded (no treaty) Indigenous territory, and wishes to salute those who, since time immemorial, have been its traditional custodians. The University expresses its respect for the contribution of Indigenous peoples to the culture of societies here and around the world. The Université de Montréal is located where, long before French settlement, various Indigenous peoples interacted with one another. We wish to pay tribute to these Indigenous peoples, to their descendants, and to the spirit of fraternity that presided over the signing in 1701 of the Great Peace of Montréal, a peace treaty founding lasting peaceful relations between France, its Indigenous allies and the Haudenosauni Con-

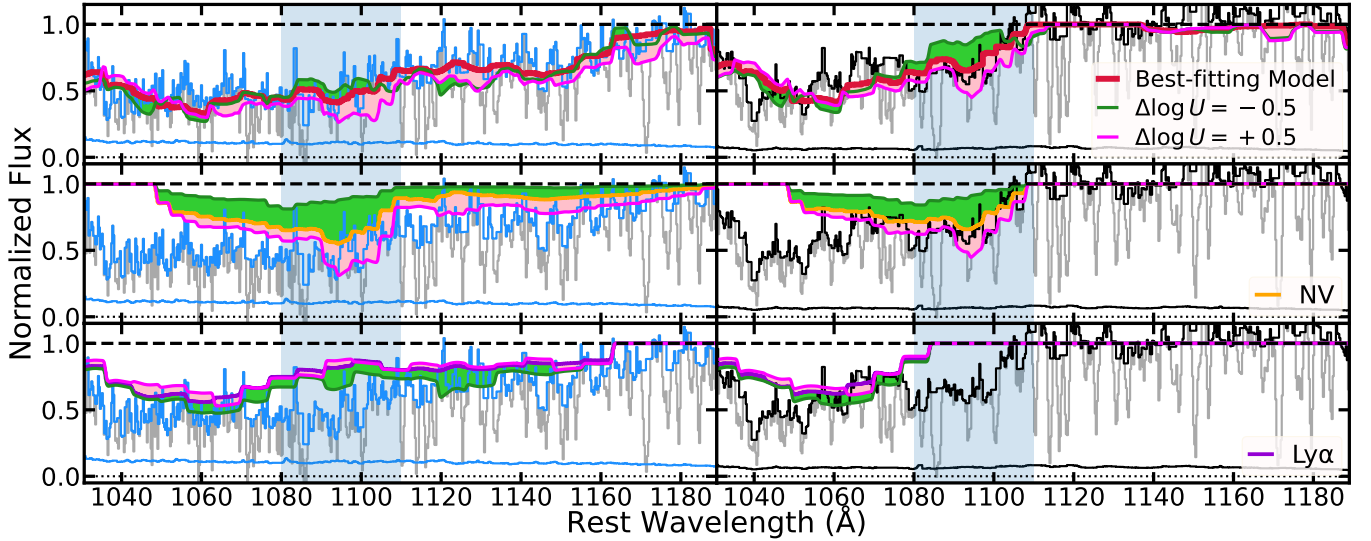

**Figure 11**: Top row shows how the best-fitting *SimBAL* models (red) change with varying ionization parameter ($\Delta \log U = \pm 0.5$; green and pink), and the two bottom rows show the change in absorption strengths of Ly$\alpha$ and N v BALs with changing $\log U$. We changed the ionization parameter by $\pm 0.5$ dex from the best-fitting values, fixed it, then refit the data to obtain the comparison *SimBAL* models. The spectra and their associated uncertainties for DR5 and DR9 are plotted in blue and black, respectively, in the left and right columns. The largest spectral variations occur around $\sim 1100$Å (blue-shaded regions), primarily driven by the substantial changes in the optical depth of N v, which scales with $\log U$ (middle row). Pixels affected by the Ly$\alpha$ forest, which have been masked (Appendix A), are shown in gray.

## APPENDIX

### A. IDENTIFICATION OF LY$\alpha$ FOREST ABSORPTION

In order to be able to analyze the potential EHVO absorption present in the Ly$\alpha$ forest, we removed intervening Ly$\alpha$ absorption lines using the following method. We used an iterative sigma-clipping method to identify the pixels affected by the Ly$\alpha$ lines. Each iteration began with a smoothed spectrum made with a Gaussian kernel of a set width, followed by flagging all data points that fell a certain sigma (e.g., $0.5\sigma \sim 2.5\sigma$) below the smoothed spectrum. The subsequent iteration produced a new smoothed spectrum with the same Gaussian kernel but with an updated spectrum where the fluxes for the flagged data points from the previous iteration had been replaced by the interpolated values from the unflagged remaining data points. The iteration ended when no new data points were flagged. We experimented with Gaussian kernels of different widths and varying sigma cutoffs, and visually inspected the results to determine the best combination for our data. We flagged 49%/39% of the data points between 1030 Å and 1216 Å as affected by non-BAL absorption following our iterative method for the SDSS/BOSS spectrum. We tested different non-BAL absorption flags produced by various sigma cutoff values for the *SimBAL* fitting and found that the solutions from the best-fitting *SimBAL* models show no significant difference.

The use of this method is possible for BAL quasar spectra because the intrinsic absorption features from the outflows are broad and smooth, unlike the narrow absorption lines from the Ly$\alpha$ forest. Otherwise, the iterative sigma-clipping method would also flag the intrinsic absorption features along with the Ly$\alpha$ lines. We do not analyze the wavelengths shortward of 1030 Å because the sensitivity and SNR are reduced at the edge of the SDSS/BOSS spectrum, and an additional opacity from Ly$\beta$ absorption lines starts to contaminate the spectra, making it more difficult to estimate the quasar continuum.

## B. CONSTRAINT OF IONIZATION PARAMETER FROM SIMBAL MODELING

In this appendix, we assess the impact of varying ionization parameter ($\log U$) on the best-fitting *SimBAL* model and demonstrate the robustness of the derived constraint. We conducted an experiment in which we varied the ionization parameter by $\pm 0.5$ dex from the best-fitting parameters, fixed the ionization parameter, and then refit the spectrum. The results from this procedure are plotted in Figure 11, which shows how the best-fitting model evolves with these varies with ionization parameter. The most significant spectral variations occur in the N v/Ly$\alpha$ BAL complex. The strength of the N v absorption scales dramatically with the ionization parameter, resulting in an over- (under-) prediction of the opacity near $\sim 1100$ Å when the ionization parameter is increased (decreased) from the best-fitting value. In contrast, the strength of the Ly$\alpha$ BAL responds to changes in the ionization parameter in the opposite direction and in a more subtle manner, with shallower absorption as the ionization parameter increases and slightly deeper absorption as it decreases. As a result, the N v transition predominantly controls the depth of the blended N v/Ly$\alpha$ BAL feature, providing a sensitive diagnostic that *SimBAL* leverages to robustly constrain the ionization parameter. While *SimBAL* performs exceptionally well in modeling heavily blended and saturated BAL troughs (Figure 6; e.g., Choi et al. 2020, 2022b), and we have effectively flagged the pixels affected by the Ly$\alpha$ forest (Appendix A), additional uncertainties in the derived outflow parameters may persist due to the N v/Ly$\alpha$ complex, which lies within this region. Future work involving a systematic analysis of EHVOs may further address and possibly clarify these issues (Rodríguez Hidalgo in prep).